\documentclass[reqno,final]{amsart}

\usepackage{caption}
\usepackage{subcaption}
\usepackage{amsfonts,latexsym,graphicx,amsmath,amssymb,bbm,enumitem,url,amsthm,color}
\usepackage{algorithm}
\usepackage{bm}
\usepackage{algpseudocode}

\newcommand{\x}{\mathbf{x}}
\newcommand{\mtensor}{\bm{\mathcal{M}}}
\newcommand{\mfield}{\mathbf{B}}
\newcommand{\etensor}{\mathbf{E}}
\newcommand{\Flux}{\bm{\Gamma}}
\newcommand{\ion}{\mathrm{i}}
\newcommand{\elec}{\mathrm{e}}
\newcommand{\ptc}{\mathrm{p}}
\newcommand{\gas}{\mathrm{g}}
\newcommand{\norm}[2]{\| #1 \|_{#2}}
\newcommand{\edges}{{\mathcal E}} 
\newcommand{\Lam}{\mathbf{\Lambda}}	
\newcommand{\V}{\mathbf{V}}
\newcommand{\disc}{{\mathcal D}}
\newcommand{\edge}{\sigma}
\newcommand{\cv}{K}
\newcommand{\grad}{\nabla}
\def\ograd{\overline{\grad}}
\newcommand{\dcvedge}{d_{\cv,\edge}}
\newcommand{\edgescv}{{{\edges}_\cv}}  
\newcommand{\centercv}{\x_\cv}   
\newcommand{\bfn}{\mathbf{n}}
\newcommand{\ncvedge}{\bfn_{\cv,\edge}}
\newcommand{\centeredge}{\overline{\mathbf{x}}_\edge}

\newtheorem{theorem}{Theorem}[section]
\newtheorem{remark}[theorem]{Remark}
\newtheorem{example}[theorem]{Example}
\begin{document}
	
	\title[]{Combining the hybrid mimetic mixed method with the Scharfetter-Gummel scheme for magnetised transport in plasmas}
	
	\author{Hanz Martin Cheng}
	\address{School of Engineering Science, Lappeenranta--Lahti University of Technology, P.O. Box 20, 53851 Lappeenranta, Finland.
		\texttt{hanz.cheng@lut.fi}}
	
	\author{Jan ten Thije Boonkkamp}
	\address{Department of Mathematics and Computer Science, Eindhoven University of Technology, P.O. Box 513, 5600 MB Eindhoven, The Netherlands.	\texttt{j.h.m.tenthijeboonkkamp@tue.nl}}
	
	\author{Jesper Janssen}
	\address{Plasma Matters B.V., Eindhoven, The Netherlands.	\texttt{jesper@plasma-matters.nl}}
	
	\author{Diana Mihailova}
	\address{Plasma Matters B.V., Eindhoven, The Netherlands.	\texttt{diana@plasma-matters.nl}}
	
	\author{Jan van Dijk}
	\address{Department of Applied Physics, Eindhoven University of Technology, P.O. Box 513, 5600 MB Eindhoven, The Netherlands.	\texttt{j.v.dijk@tue.nl}}
	
	\vspace{10pt}
	
	\date{\today}
	
	\begin{abstract}
		In this paper, we propose a numerical scheme for fluid models of magnetised plasmas.  One important feature of the numerical scheme is that it should be able to handle the anisotropy induced by the magnetic field. In order to do so, we propose the use of the hybrid mimetic mixed (HMM) scheme for diffusion. This is combined with a hybridised variant of the Scharfetter-Gummel (SG) scheme for advection. The proposed hybrid scheme can be implemented very efficiently via static condensation. Numerical tests are then performed to show the applicability of the combined HMM-SG scheme, even for highly anisotropic magnetic fields.
	\end{abstract}
	
	\maketitle
	
	%
	%
	%
	%
	%

	\section{Introduction}
	\noindent Magnetostatic fields play an important role in plasma physics. One useful property of a magnetic field is that it confines the plasma, limiting charged particle loss to the walls; hence, allowing magnetron discharges to operate at relatively low voltages. Some examples of these can be seen in  Hall-effect thrusters (HETs) and electron-cyclotron-resonance (ECR)
	discharges \cite{H07-model}. These physical processes are expensive to set-up, and so we turn to mathematical models and perform numerical simulations in order to get an overview and general understanding of how particles behave in these situations.
	
	Early numerical studies for plasmas involve particle-based models, which involve Monte Carlo simulations \cite{BL85-particleBook,D64-particle}. However, the inclusion of a magnetic field complicates the physics and hence the numerical modeling of magnetised discharges. In this paper, we consider fluid models. These consist of drift-diffusion equations for the particles (electrons and ions), coupled to the Poisson equation for the electric field, see e.g.: \cite{C05-fluid_ion,H07-model,Hetal02-thruster,K03-fluid_ion,P94-model}.  This is an approximation for the real physics which is computationally more efficient and avoids some of the numerical complications and high computational costs associated to using Monte Carlo type methods on purely particle-based models \cite{H07-model}.  
	
	The model system of equations is highly nonlinear. Moreover, the Poisson equation and the drift-diffusion equations are strongly coupled for the computation of the electric field and the charge densities. In order to resolve this, we decouple the equations by using the transversal method of lines, where a semi-discrete system is obtained after discretising with respect to the time variable. Following this, we look at the spatial discretisation. The Poisson equation is elliptic, and can thus be solved by classical techniques, such as the central difference method. In the absence of a magnetic field, the drift-diffusion equations can be solved efficiently and precisely by the Scharfetter-Gummel (SG) scheme \cite{SG69}. On the other hand, the presence of a magnetic field introduces anisotropy in the drift-diffusion equations, hence requiring a numerical scheme which captures the anisotropic diffusion accurately. Here, we only explore linear diffusion tensors. Nonlinear diffusion tensors can be addressed by considering a generalisation of the SG scheme \cite{BC17-exp_decay,CCFG20-FV}, but are not in the scope of this paper. Recent studies propose the use of high order finite difference gradient reconstruction methods, together with a magnetic field aligned mesh \cite{PGO16-anisotropy,ZPFA19-numerics} in order to resolve the anisotropic diffusion. 
	
	The aim of this paper is to propose a numerical scheme which is computationally efficient, and is able to handle the anisotropy induced by the magnetic field. In order to do so, we propose, as in \cite{VDM11-adv-diff}, to separate the discretisation of the diffusive and advective fluxes. We then use the hybrid-mimetic-mixed (HMM) method \cite{dro-10-uni} for discretising the diffusive flux, and a modification of the SG scheme for the advective flux. The main interest of considering the HMM method is its ability to handle the anisotropic diffusion tensor on generic meshes.  Moreover, the method can be fully hybridised, leading to a purely local stencil for solving the system of equations, which allows a very efficient implementation of the scheme. Also, the combination of the HMM with the SG scheme has been widely explored, and has several desirable properties for drift-diffusion equations, see e.g. \cite{CD09-FV,CHLM22-longtermFV}.  As a remark, the HMM method can also be viewed as a gradient reconstruction method or a gradient scheme \cite{GDMBook16}, and can thus be used in combination with the methods proposed in \cite{PGO16-anisotropy,ZPFA19-numerics}. 
	
	The outline of this paper is as follows. We start by presenting the mathematical model for magnetised transport in Section \ref{sec:MathModel}. This model consists of a set of drift-diffusion equations, coupled to the Poisson equation. We then describe in Section \ref{sec:Time_disc} the choice of time discretisation, and how to decouple the system. Following this, we employ a finite volume type spatial discretisation in Section \ref{sec:Space_disc}, using the HMM for the diffusive fluxes, and a modified SG for the advective fluxes. Details of the implementation are then given in Section \ref{sec:Imp}. Numerical tests are provided in Section \ref{sec:Numtests} to show the effectiveness of the proposed method. Finally, in Section \ref{sec:Conc} we provide a summary of the results and some recommendations for future research.
	
	\section{Mathematical model} \label{sec:MathModel}
	\noindent 
	Suppose that an electric field $\etensor$ is applied between two parallel electrodes and that an electron 
	flux is forced through a uniform medium, where an oblique magnetic field $\mfield$ is applied. The electric field $\etensor=-\nabla V$  in a spatial domain $\Omega$ over a time interval $(0,T)$ is governed by the Poisson equation 
	\begin{subequations}
		\begin{equation}\label{eq:Poisson}
			\nabla \cdot (\varepsilon\etensor) = \rho \qquad \mathrm{on} \quad \Omega\times(0,T),
		\end{equation}
		where $\varepsilon$ is the permittivity of vacuum, $V$ is the potential, and  $\rho$ is the space charge density given by 
		\begin{equation}\nonumber
			\rho = q(n_\mathrm{i}-n_\mathrm{e}).
		\end{equation}
		Here, $q$ is the elementary charge,  $n_\ion$ and $n_\elec$ are the ion and electron densities, respectively.
		Under the assumption that ions are not magnetised, the model which describes the evolution of the ion and electron densities is then given by the drift-diffusion equations:
		
		\begin{align}
			\frac{\partial n_\ion}{\partial t} + \nabla \cdot \Flux_\ion -k n_{\mathrm{e,loc}}n_\gas= 0 \qquad \mathrm{on} \quad \Omega \times(0,T),\label{eq:magnetised_transport_ion}\\
			\frac{\partial n_\elec}{\partial t} + \nabla \cdot \mtensor_\elec\Flux_\elec -kn_{\mathrm{e,loc}}n_\gas= 0 \qquad \mathrm{on} \quad \Omega \times(0,T),\label{eq:magnetised_transport_electron}
		\end{align}
		where $\Flux_\ion$ and $\Flux_\elec$ are the drift-diffusion flux densities for the ions and electrons, respectively, defined by
		\begin{equation}\nonumber
			\begin{aligned}
				\Flux_\ion &:= \mu_\ion \etensor n_\ion - D_\ion\nabla n_\ion,\\
				\Flux_\elec &:=  \mu_\elec \etensor n_\elec -  D_\elec \nabla n_\elec.
			\end{aligned}
		\end{equation}
		Equations \eqref{eq:magnetised_transport_ion} and \eqref{eq:magnetised_transport_electron} are coupled to the Poisson equation \eqref{eq:Poisson} due to the electric field $\etensor$ and the space charge density $\rho$. 
	\end{subequations}
	The parameters and constants used in the model are enumerated in Tables \ref{tab:param_model} and \ref{tab:const_model}. For quantities which are common to both ions and electrons, we use the subscript $\ptc = \ion,\elec$, referring to a particle (either an ion or an electron).
	\begin{table}[h!]
		\caption{Parameters in the model.}\label{tab:param_model}
		\centering
		\begin{tabular}{|cc|}
			\hline
			$k$  & ionisation rate \\
			$\mu_\ptc$   & mobility coefficient \\	
			$D_\ptc$ & diffusion coefficient   \\	
			$\mtensor_\elec$ & magnetic tensor \\ 	
			$n_{\mathrm{e,loc}}$ & local electron density\\
			\hline
		\end{tabular} 
	\end{table}
	\begin{table}[h!]
		\caption{Constants in the model.}\label{tab:const_model}
		\centering
		\begin{tabular}{|cc|}
			\hline
			$n_\gas$ & 	density of the background gas   \\
			$\bar{\epsilon}=2$eV & mean electron energy \\
			$\varepsilon$   & permittivity of vacuum \\
			$k_\mathrm{B}$  & Boltzmann constant\\
			$q$ & elementary charge\\
			$T_\gas$ & temperature of the background gas\\
			\hline
		\end{tabular} 
	\end{table}

\begin{figure}[h]
		\centering
		\includegraphics[width=0.65\linewidth]{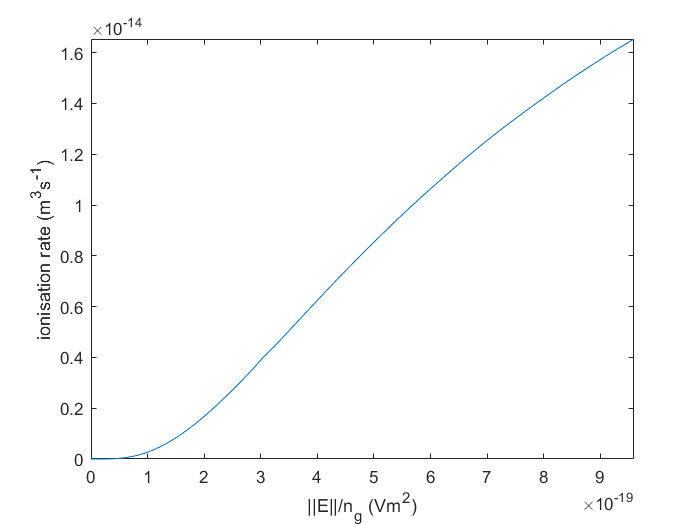}
		\caption{Ionisation rate $k$} \label{fig.lookup_k}
	\end{figure}

\begin{figure}[h]
		\begin{tabular}{cc}
			\includegraphics[width=0.5\linewidth]{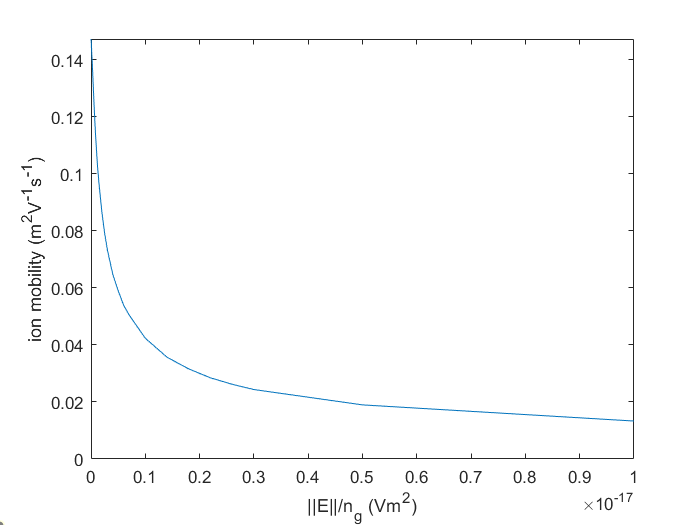} &
			\includegraphics[width=0.5\linewidth]{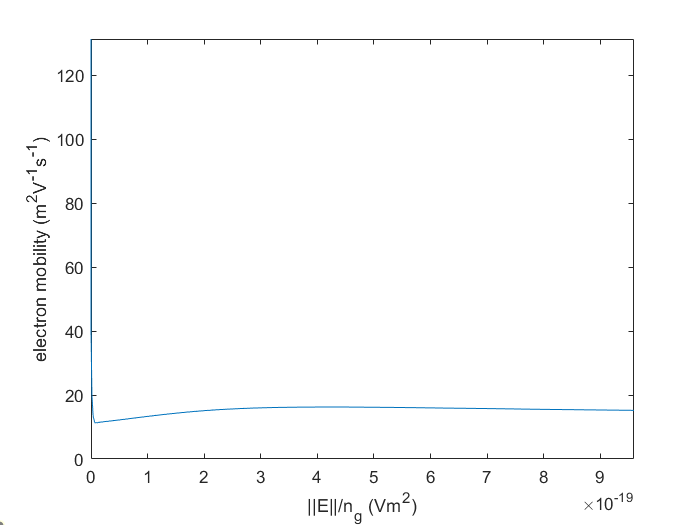} 
		\end{tabular}
		\caption{Mobility coefficients (Left: $\mu_\ion$, Right:$\mu_\elec$ ).} \label{fig.lookup_mu}
	\end{figure}

	We now discuss the parameters used in the model. Firstly, the ionisation rate $k$ and the mobility coefficients $\mu_\ptc$ are computed via local field approximations, and are hence functions of the reduced electric field $\norm{\etensor}{} /n_\gas$. In this paper, we consider helium gas, whose ion mobility $\mu_\ion$ (Figure \ref{fig.lookup_mu}, left) is obtained from \cite{E76-ion_mob}. The ionisation rate $k$ (Figure \ref{fig.lookup_k}) and electron mobility $\mu_\elec$ (Figure \ref{fig.lookup_mu}, right) were computed in \cite{H05_neLoc} using cross-sectional data for helium gas in \cite{CER70-cross,MC77-cross} for low energies and \cite{H81-cross} for high energies. Here and throughout the paper, $\norm{\cdot}{}$ refers to the $L^2$-norm. Secondly, we discuss the diffusion coefficients. These are governed by the Stokes-Einstein relation 
	
	\begin{equation}\nonumber
		D_\ptc = \frac{k_\mathrm{B} T_\ptc \mu_\ptc}{q_\ptc},
	\end{equation} where, $q_\ion = q$ and $q_\elec = -q$.  For ions, we assume that $T_\ion = T_\gas$ and so we have
	\begin{subequations}
		\begin{equation}\label{eq:Di_ion}
			D_\ion = \frac{k_\mathrm{B}T_\gas\mu_\ion}{q}.
		\end{equation}
		We assume that electrons follow a Maxwellian distribution so that $k_\mathrm{B} T_\elec = 2\bar{\epsilon} /3 $ and thus
		\begin{equation} \label{eq:Di_elec}
			D_\elec = -\bar{\epsilon} \frac{2\mu_\elec}{3q}. 
		\end{equation} 
	\end{subequations}
	Thirdly, the local electron density $n_{\mathrm{e, loc}}$ is computed by the relation \cite{H05_neLoc}
	\begin{equation}\label{eq:loc_elec_dens}
		n_{\mathrm{e,loc}} = \frac{\norm{ \mtensor_\elec\Flux_\elec}{}}{\mu_\elec\norm{ \mtensor_\elec \etensor }{}}.
	\end{equation}
	Finally, the magnetic tensor $\mtensor_\elec$ is given by
	\begin{subequations}
		\begin{align}
			\mtensor_\elec &= \frac{1}{\norm{\bm{\beta}}{}^2+1} \begin{bmatrix} 
				1+\beta_1^2 & \beta_3+\beta_2\beta_1 & -\beta_2+\beta_3\beta_1 \\
				-\beta_3+\beta_2\beta_1 & 1+\beta_2^2 & \beta_1+\beta_3\beta_2 \\
				\beta_2+\beta_3\beta_1 & -\beta_1+\beta_3\beta_2 & 1+\beta_3^2 
			\end{bmatrix} \nonumber \\
			&= \frac{1}{\norm{\bm{\beta}}{}^2+1} \big(\mathbf{I}+\bm{\beta}\bm{\beta}^T 
			-\mathbf{C}(\bm{\beta})\big)\label{def:Mtensor},\\
			\mathbf{C}(\bm{\beta}) &= \begin{bmatrix}
				0 & -\beta_3 & \beta_2 \\
				\beta_3 & 0 &  	-\beta_1 \\
				-\beta_2 & \beta_1 & 0 
			\end{bmatrix},\nonumber
		\end{align}
		with Hall parameters $\beta_i = \mu_\elec B_i$, $i=1,2,3$ and magnetic field $\mathbf{B}$. We now note that $\mathbf{C}(\bm{\beta})$ is a cross product matrix such that for any vector $\mathbf{v}\in\mathbb{R}^3$, $\mathbf{C}(\bm{\beta}) \mathbf{v} = \bm{\beta}\times \mathbf{v}$. At this stage, it is also important to note that the magnetic tensor $\mtensor_\elec$ is positive semidefinite. That is, $\mathbf{v}^T \mtensor_\elec \mathbf{v} \geq 0$ for any vector $\mathbf{v}$ in $\mathbb{R}^3$. We also note that 
		\begin{equation}\nonumber
			\begin{aligned}
				\mtensor_\elec \bm{\beta} &= \frac{1}{\norm{\bm{\beta}}{}^2+1}\big(\bm{\beta} + \bm{\beta}(\bm{\beta}^T\bm{\beta})-\bm{\beta}\times \bm{\beta} \big)
				\\ &= \bm{\beta},
			\end{aligned}
		\end{equation}
		which shows that $\bm{\beta}$ is an eigenvector of $\mtensor_\elec$ corresponding to the eigenvalue 1. Essentially, this means that $\mtensor_\elec \mathbf{v}$ preserves the component of the vector $\mathbf v$ that is parallel to the magnetic field. 

			For dimension $d=2$,
			the magnetic field can be specified at an angle $\theta$ with respect to the $x$-axis by setting
			\begin{equation}\label{eq:mag_angle}
				\begin{aligned} \mathbf{B}&=\big(B_1,B_2,0\big)^T\\
					&=\norm{\mathbf{B}}{}\big(\cos\theta,\sin\theta,0\big)^T,
				\end{aligned}
			\end{equation}
		\end{subequations}
		where $\norm{\mathbf{B}}{}$ is the strength of the magnetic field. Considering now a vector
		\begin{equation}\nonumber
			\bm{\beta}^\perp = \mu_\elec \norm{\mathbf{B}}{}\big(-\sin\theta,\cos\theta,0\big)^T
		\end{equation} orthogonal to the magnetic field, we see that 
		\begin{equation}\label{eq:mag_field_perp}
			\mtensor_\elec\bm{\beta}^\perp = \frac{1}{\norm{\bm{\beta}}{}^2+1}(\bm{\beta}^\perp - \norm{\bm{\beta}}{}^2\mathbf{e}_z).
		\end{equation}For two-dimensional models, we neglect the $z$-component, and hence we see that the components orthogonal to the magnetic field are scaled by a factor $1/(\norm{\bm{\beta}}{}^2+1)$. These properties agree with the classical definition \cite{LL03-plasma} of a magnetic tensor.
		
		\subsection{Boundary conditions}
		We now describe the boundary conditions for the model. At the electrodes (corresponding to either anodes or cathodes), Dirichlet boundary conditions are imposed for the Poisson equation \eqref{eq:Poisson},
		\begin{subequations}
			\begin{equation}\label{eq:DBc}
				\begin{aligned}
					V &= V_\mathrm{a} \quad \mathrm{anode},\\
					V &= V_\mathrm{c} \quad \mathrm{cathode}.
				\end{aligned}
			\end{equation}
			Correspondingly, flux boundary conditions are imposed for the drift-diffusion equations \eqref{eq:magnetised_transport_ion} and \eqref{eq:magnetised_transport_electron} at the electrodes, i.e.,
			\begin{align}
				\label{eq:fluxBcI}\Flux_\ion \cdot \mathbf{n} &= (\mu_\ion \etensor \cdot  \mathbf{n} )^+ n_\ion + \frac{1}{4}v_{\mathrm{th, i}}n_\ion,\\
				\label{eq:fluxBcE} (\mtensor_\elec \Flux_\elec) \cdot \mathbf{n} &= (\mu_\elec \mtensor_\elec \etensor \cdot  \mathbf{n} )^+ n_\elec + \frac{1}{4}v_{\mathrm{th, e}}n_\elec - \gamma_\ion\Flux_\ion\cdot \mathbf{n}, 
			\end{align}
			where $v_{\mathrm{th, p}}$ is the thermal velocity, $f^+=\max(f,0)$ denotes the positive part of the function $f$,  and $\mathbf{n}$ is the outward unit normal vector at the boundary. The first two terms in the right hand side of equations \eqref{eq:fluxBcI} and \eqref{eq:fluxBcE} describe the loss of electrons and ions, respectively, through the wall via the drift and thermal fluxes. On the other hand, the third term in \eqref{eq:fluxBcE} is the secondary electron emission, which describes the production of electrons as a result of ion fluxes bombarding the wall \cite{B88-BC}, where $\gamma_\ion$ is the average number of electrons emitted per incident ion.

			Finally,  homogeneous Neumann boundary conditions are imposed over the other boundaries of the domain. That is, for \eqref{eq:Poisson}, \eqref{eq:magnetised_transport_ion}, and \eqref{eq:magnetised_transport_electron} we impose:
			\begin{equation}\label{eq:NBc}
				\begin{aligned}
					\nabla V \cdot \mathbf{n} &= 0, \\
					\nabla n_\ion \cdot \mathbf{n}&= 0,\\
					\nabla n_\elec \cdot \mathbf{n}&= 0.\\
				\end{aligned}
			\end{equation}
		\end{subequations}
		For the drift-diffusion equations, this means that there is no diffusion of ions and electrons across these boundaries. For the Poisson equation, this means that the electric field lines are parallel to the boundary.
		
		Given the electron and ion densities  $n_\elec(\x,0), n_\ion(\x,0)$ at time $t=0$, we are then interested in calculating the densities $n_\elec(\x,T), n_\ion(\x,T)$  at time $t=T$. In this work, we consider a numerical scheme based on the transversal method of lines. That is, we start by presenting a semi-discrete version of the model system of equations with respect to the time variable. Following this, we then describe the discretisation in space.
		
		\section{Time discretisation}\label{sec:Time_disc}
		\noindent We form a partition of the time interval $(0,T)$ by taking $0=t^{(0)}<t^{(1)}<\dots t^{(M)}=T$. A semi-implicit Euler method will be considered for the time integration. This will be obtained by starting off with a fully implicit Euler method, for which some of the implicit terms will be made explicit in order to have a linear scheme. Even though the Euler method is only first-order accurate in time, it is sufficient since we are only interested in the steady-state solution. Using a uniform time step of $\Delta t=t^{(m+1)}-t^{(m)}$ for $m=0,\dots M-1$, and given the densities $n_\elec^{(m)}\approx n_\elec(\x,t^{(m)}), n_\ion^{(m)} \approx n_\ion(\x,t^{(m)})$, we seek $\etensor^{(m+1)}, n_\elec^{(m+1)}, n_\ion^{(m+1)}$ such that 
		
		\begin{subequations}\label{eq:implicit_Euler}
			
			\begin{equation}\label{eq:disc_Poisson}
				\nabla \cdot (\varepsilon\etensor^{(m+1)}) = q(n_{\ion}^{(m+1)}-n_{\elec}^{(m+1)}),
			\end{equation}
			
			\begin{equation}\label{eq:disc_magnetised_transport_ions}
				\frac{n_\ion^{(m+1)}-n_\ion^{(m)}}{\Delta t} + \nabla \cdot \Flux_\ion^{(m+1)} -k^{(m+1)}n_{\mathrm{e, loc}}^{(m+1)}n_\gas= 0,
			\end{equation}
			
			\begin{equation}\label{eq:disc_magnetised_transport_electrons}
				\frac{n_\elec^{(m+1)}-n_\elec^{(m)}}{\Delta t} + \nabla \cdot \mtensor_e^{(m+1)} \Flux_\elec^{(m+1)} -k^{(m+1)}n_{\mathrm{e, loc}}^{(m+1)}n_\gas= 0.
			\end{equation}
		\end{subequations}
		Here,
		\begin{equation}\nonumber
			\Flux_\ptc^{(m+1)}= \mu_\ptc^{(m+1)} \etensor^{(m+1)}n_\ptc^{(m+1)}  - D_\ptc^{(m+1)}\nabla n_\ptc^{(m+1)}
		\end{equation}
		is the drift-diffusion flux density of particle $\ptc$ (ions or electrons) at time $t^{(m+1)}$. As a remark, we note that even though the scheme was presented with a uniform time stepping, it can be easily adapted into a scheme with an adaptive time stepping $\Delta t^{(m)} = t^{(m+1)}-t^{(m)}$.
		
		We note that due to the electric field, the Poisson equation \eqref{eq:disc_Poisson} is coupled to the continuity equations \eqref{eq:disc_magnetised_transport_ions} and \eqref{eq:disc_magnetised_transport_electrons}. This results in a nonlinear scheme, which is computationally very expensive. To avoid this, we consider instead a semi-implicit scheme which is obtained by decoupling the system of equations \eqref{eq:disc_Poisson}-\eqref{eq:disc_magnetised_transport_electrons}.
		
		\subsection{First-order approximation of the charge density}
		The simplest way to decouple the Poisson equation from the continuity equations is to use a first-order approximation of the charge density. This leads to seeking, for $m=0,\dots M-1$, $\etensor^{(m+1)}$ such that
		
		\begin{equation}\label{eq:disc_Poisson_n}
			\nabla \cdot (\varepsilon\etensor^{(m+1)}) = q(n_{\ion}^{(m)}-n_{\elec}^{(m)}).
		\end{equation}
		The advantage of using \eqref{eq:disc_Poisson_n} is that the charge density will then only depend on the electron and ion densities at the previous time step. On the other hand, this is a first-order approximation, which means that the leading order of error in time is $O(\Delta t)$. Hence, a very small time step, at most equal to the dielectric relaxation time \cite[Section II.C.]{V94-Ventzek1}, is needed in order to ensure that the electric field does not change sign during one time step, and that the approximation of the electric field is good enough to be used for computing the electron and ion densities in the drift-diffusion equations.

		\subsection{Second-order approximation of the charge density}
		
		We now present a second-order approximation by performing a Taylor expansion on the charge density.
		%
		%
		Centering the Taylor expansion at time $t^{(m+1)}$, we may write
		\begin{equation}\label{eq:charge_density_Taylor_n1}
			\rho(\x,t^{(m+1)}) = \rho(\x,t^{(m)}) + \Delta t \frac{\partial \rho(\x,t^{(m+1)})}{\partial t}+ O(\Delta t^2). 
		\end{equation} 
		We then note that the time derivative $\frac{\partial \rho}{\partial t}$ can be computed from the drift-diffusion equations, and can be written as
		\begin{equation}\label{eq:drho_dt}
			\frac{\partial \rho}{\partial t} = -q\big(\nabla \cdot \Flux_\ion - \nabla \cdot \mtensor_\elec \Flux_\elec \big). 
		\end{equation}
		Using now \eqref{eq:charge_density_Taylor_n1} for the charge density in the Poisson equation and evaluating $\frac{\partial \rho}{\partial t}$ in \eqref{eq:drho_dt} at time $t^{(m+1)}$, we then seek, for $m=0,\dots M-1$, $\etensor^{(m+1)}$ such that
		\begin{equation}\label{eq:disc_Poisson_n1_imp}
			\begin{aligned}
				\nabla \cdot (\varepsilon\etensor^{(m+1)}) = &  q\big(n_{\ion}^{(m)}-n_{\elec}^{(m)}\big)-q\Delta t\nabla \cdot \Flux_\ion^{(m+1)}\\
				&+q\Delta t\nabla \cdot \big(\mtensor_\elec^{(m+1)} \Flux_\elec^{(m+1)} \big).
			\end{aligned}
		\end{equation}
	
		In this case, the quantities $\etensor,\mtensor_\elec, \mu_\ion,\mu_\elec,D_\ion,D_\elec,n_\ion,$ and $n_\elec$ at time $t^{(m+1)}$ are unknown and hence the Poisson equation is still coupled to the drift-diffusion equations. One way to deal with this is to consider a semi-implicit treatment of these quantities \cite{V94-Ventzek1,V93-Ventzek2}. That is, we keep $\etensor$ implicit and take first-order approximations to $\mtensor_\elec,\mu_\ion,\mu_\elec,D_\ion,D_\elec,n_\ion,n_\elec$, which leads to
		\begin{equation}\nonumber
			\begin{aligned}
				\nabla \cdot (\varepsilon\etensor^{(m+1)}) = &  q\big(n_{\ion}^{(m)}-n_{\elec}^{(m)}\big)-q\Delta t\nabla \cdot \bigg(\mu_\ion^{(m)}\etensor^{(m+1)} n_\ion^{(m)}  -D_\ion^{(m)}\nabla n_\ion^{(m)}\bigg)\\
				&+q\Delta t\nabla \cdot \mtensor_\elec^{(m)} \bigg(\mu_\elec^{(m)}\etensor^{(m+1)} n_\elec^{(m)}  -D_\elec^{(m)}\nabla n_\elec^{(m)}\bigg),
			\end{aligned}
		\end{equation}
		or equivalently,
		\begin{equation}\nonumber
			\begin{aligned}
				\nabla \cdot \bigg(\bigg(\varepsilon\mathbf{I}+q\Delta t&\big(\mu_\ion^{(m)}n_\ion^{(m)}\mathbf{I}-\mu_\elec^{(m)}n_\elec^{(m)}\mtensor_\elec^{(m)}\big)\bigg)\etensor^{(m+1)}\bigg)\\
				&\!\!\!  =q(n_{\ion}^{(m)}-n_{\elec}^{(m)})+q\Delta t\nabla \cdot \big(D_\ion^{(m)}\nabla n_\ion^{(m)}\big)\\
				&-q\Delta t\nabla \cdot \big(\mtensor_\elec^{(m)} D_\elec^{(m)}\nabla n_\elec^{(m)}\big) .
			\end{aligned}
		\end{equation}
		
		\noindent Denoting the correction flux at time level $m+1$ by
		\begin{equation}\label{eq:corr_flux}
			\Flux_\mathrm{c}^{(m+1)} := \big(\mu_\ion^{(m)}n_\ion^{(m)}\mathbf{I}-\mu_\elec^{(m)}n_\elec^{(m)}\mtensor_\elec^{(m)}\big)\etensor^{(m+1)},
		\end{equation}
		the above equation can be rewritten as 
		\begin{equation}\label{eq:disc_Poisson_n1_semi_imp}
			\begin{aligned}
				\nabla \cdot (\varepsilon\etensor^{(m+1)}) +q\Delta t \nabla \cdot\Flux_\mathrm{c}^{(m+1)}
				&\!  =q(n_{\ion}^{(m)}-n_{\elec}^{(m)})+q\Delta t\nabla \cdot \big(D_\ion^{(m)}\nabla n_\ion^{(m)}\big)\\
				&-q\Delta t\nabla \cdot \big(\mtensor_\elec^{(m)} D_\elec^{(m)}\nabla n_\elec^{(m)}\big) .
			\end{aligned}
		\end{equation}
		
		We now look at the coercivity properties of the modified Poisson equation which results from the second-order approximation of the charge density. Due to the correction terms, the matrix $\mathbf{P} = \varepsilon \mathbf{I}$ associated to the Poisson equation is now modified and we have $ \mathbf{\widehat{P}} = \varepsilon\mathbf{I}+q\Delta t \big(\mu_\ion^{(m)}n_\ion^{(m)}\mathbf{I}-\mu_\elec^{(m)}n_\elec^{(m)}\mtensor_\elec^{(m)}\big)$. We now note that $\mu_\ion^{(m)} n_\ion^{(m)} >0$ and $\mu_\elec^{(m)} n_\elec^{(m)} <0$. Moreover, $\mtensor_\elec^{(m)}$ is positive semidefinite, and thus we have that $\mathbf{v}^T\mathbf{\widehat{P}}\mathbf{v}\geq \varepsilon   \mathbf{v}^T\mathbf{v}$ for any vector $\mathbf{v}$, which ensures the coercivity of the matrix $\mathbf{\widehat{P}}$ associated to the modified Poisson equation. 
		\section{Spatial discretisation}\label{sec:Space_disc}
		\noindent For the spatial discretisation, we start by forming $\Omega_h$, a partition of the domain $\Omega$ into $N_K$ control volumes. Each of the control volumes $K$ has edges $\sigma \in \edges_K$ such that $\max_{\sigma\in\edges_K,K\in\Omega_h} |\sigma| =h$, and we denote by $N_\sigma$ the total number of edges contained in this partition. We then denote by $X_\disc$ the space which contains the collection of unknowns (one on each cell, and one on each edge). For this work, we focus on finite volume type schemes.
		
		\subsection{Finite volume discretisation}\label{sec:FV}
		
		To write the equations \eqref{eq:disc_Poisson},\eqref{eq:disc_magnetised_transport_ions}, and \eqref{eq:disc_magnetised_transport_electrons} in their finite volume forms, we take their integrals over all control volumes $K\in\Omega_h$. This results in, for each control volume $K$,

		\begin{equation}\nonumber
			\int_K \nabla \cdot (\varepsilon\etensor^{(m+1)}) \mathrm{d}A= q\int_K (n_{\ion}^{(m+1)}-n_{\elec}^{(m+1)}) \mathrm{d}A,
		\end{equation}
		
		\begin{equation}\nonumber
			\frac{1}{\Delta t}\int_K \big(n_\ion^{(m+1)}-n_\ion^{(m)}\big) \mathrm{d}A + \int_K  \nabla \cdot \Flux_\ion^{(m+1)} \mathrm{d}A -n_\gas\int_K k^{(m+1)}n_{\mathrm{e, loc}}^{(m+1)} \mathrm{d}A= 0,
		\end{equation}
		
		\begin{equation}\nonumber
			\frac{1}{\Delta t}\int_K \big(n_\elec^{(m+1)}-n_\elec^{(m)}\big) \mathrm{d}A+ \int_K \nabla \cdot \mtensor_\elec^{(m+1)}\Flux_\elec^{(m+1)} \mathrm{d}A - n_\gas\int_K k^{(m+1)}n_{\mathrm{e, loc}}^{(m+1)} \mathrm{d}A= 0,
		\end{equation}
					respectively. By applying the divergence theorem and using $\etensor^{(m+1)}=-\nabla V^{(m+1)}$,
					these can be rewritten as
					\begin{subequations}
						\begin{equation}\label{eq:FV_Poisson}
							-\sum_{\sigma\in\edges_K} \int_\sigma \varepsilon \nabla V^{(m+1)} \cdot \mathbf{n}_{K,\sigma} \mathrm{d} s=q\int_K (n_\ion^{(m+1)}-n_\elec^{(m+1)}) \mathrm{d} A,
						\end{equation}	
						\begin{equation}\label{eq:FV_ion}
							\begin{aligned}
								\frac{1}{\Delta t}\int_K \big(n_\ion^{(m+1)}-n_\ion^{(m)}\big) \mathrm{d}A &+ \sum_{\sigma\in\edges_K} \int_\sigma  \Flux_\ion^{(m+1)} \cdot \mathbf{n}_{K,\sigma} \mathrm{d} s  \\
								& - n_\gas\int_K k^{(m+1)} n_{\mathrm{e, loc}}^{(m+1)} \mathrm{d} A=0,
							\end{aligned}
						\end{equation}
						\begin{equation}\label{eq:FV_electron}
							\begin{aligned}
								\frac{1}{\Delta t} \int_K \big(n_\elec^{(m+1)}-n_\elec^{(m)}\big) \mathrm{d}A &+ \sum_{\sigma\in\edges_K} \int_\sigma \mtensor_\elec^{(m+1)} \Flux_\elec^{(m+1)} \cdot \mathbf{n}_{K,\sigma} \mathrm{d} s \\
								& -n_\gas\int_K k^{(m+1)} n_{\mathrm{e, loc}}^{(m+1)}  \mathrm{d} A=0.
							\end{aligned}
						\end{equation}	
					\end{subequations}
					Here, $\mathbf{n}_{K,\sigma}$ is the outward unit normal vector of $K$ at $\sigma$. Due to the definition \eqref{eq:loc_elec_dens} of $n_{\mathrm{e, loc}}$, the term $n_{\mathrm{e, loc}}^{(m+1)}$ is nonlinear in $n_\elec$. Here, we choose a first-order approximation and take the local electron density at time $m$ so that
					\begin{equation}\label{eq:disc_ne_loc}
						n_{\mathrm{e, loc}}^{(m)}:=  \frac{\norm{ \mtensor_\elec^{(m)}\Flux_\elec^{(m)} }{}}{\mu_\elec^{(m)}\norm{ \mtensor_\elec^{(m)} \etensor^{(m)} }{}}.
					\end{equation} 
					
					Key to the definition of a finite volume scheme is the choice of how to discretise the fluxes. In order to be able to treat the anisotropy, we propose to split the fluxes into a diffusive and an advective component. Diffusive fluxes are of the form $-\int_\sigma \Lam \nabla c \cdot \mathbf{n}_{K,\sigma} \mathrm{d} s$ for some diffusion tensor $\Lam$ and scalar $c$; on the other hand, the advective fluxes take the form $\int_\sigma  c \V \cdot \mathbf{n}_{K,\sigma} \mathrm{d} s$, where $\V$ is a velocity field. In equations \eqref{eq:FV_Poisson}-\eqref{eq:FV_electron}, three diffusive fluxes and two advective fluxes are present. 
					We then denote the discrete diffusive and advective fluxes for each edge $\sigma$ in a cell $K\in \Omega_h$ by
					\begin{subequations}\label{not:fluxes}
						\begin{equation}\label{not:diff_fluxV}
							F_{K,\sigma}^{D_V} \approx -\int_\sigma \varepsilon \nabla V \cdot \mathbf{n}_{K,\sigma} \mathrm{d} s,
						\end{equation}
						\begin{equation}\label{not:diff_fluxi}
							F_{K,\sigma}^{D_\ion} \approx -\int_\sigma D_\ion \nabla n_\ion \cdot \mathbf{n}_{K,\sigma}\mathrm{d} s,
						\end{equation}
						\begin{equation}\label{not:diff_fluxe}
							F_{K,\sigma}^{D_\elec} \approx -\int_\sigma \mtensor_\elec D_\elec\nabla n_\elec \cdot \mathbf{n}_{K,\sigma}\mathrm{d} s,
						\end{equation}
						\begin{equation}\label{not:adv_fluxi}
							F_{K,\sigma}^{A_\ion} \approx \int_\sigma n_\ion\mu_\ion\etensor \cdot \mathbf{n}_{K,\sigma}\mathrm{d} s,
						\end{equation}
						\begin{equation}\label{not:adv_fluxe}
							F_{K,\sigma}^{A_\elec} \approx \int_\sigma n_\elec\mu_\elec\mtensor_\elec \etensor \cdot \mathbf{n}_{K,\sigma}\mathrm{d} s.
						\end{equation}
						
					\end{subequations}
					Replacing the fluxes in \eqref{eq:FV_Poisson}-\eqref{eq:FV_electron} with their discrete counterparts \eqref{not:diff_fluxV}-\eqref{not:adv_fluxe} at the appropriate time level and under the assumption that $n_{\ion}^{(m)}$ and $n_\elec^{(m)}$ are piecewise constant with values $n_{\ion,K}^{(m)}$ and $n_{\elec,K}^{(m)}$ respectively in a cell $K$, we then obtain our numerical scheme. For the Poisson equation, we start by considering a first-order approximation of the charge density, given in \eqref{eq:disc_Poisson_n}. That is, for $m=0,\dots, M$, given the densities $n_\elec^{(m)}\approx n_\elec(\x,t^{(m)}), n_\ion^{(m)} \approx n_\ion(\x,t^{(m)})$, we seek $\etensor^{(m+1)}, n_\elec^{(m+1)}, n_\ion^{(m+1)}$ such that for each $K\in \Omega_h$,
					\begin{subequations}\label{eq:full_scheme}
						\begin{equation}\label{eq:full_scheme_Poisson}
							\sum_{\edge \in \edgescv} F_{K,\sigma}^{D_V,(m+1)} = q |K| (n_{\ion,K}^{(m)}-n_{\elec,K}^{(m)}),
						\end{equation}
						\begin{equation}\label{eq:full_scheme_ion}
							|K| \frac{n_{\ion,K}^{(m+1)}}{\Delta t}  +\sum_{\sigma\in\edges_K} F_{K,\sigma}^{D_\ion,(m+1)} + \sum_{\sigma\in\edges_K} F_{K,\sigma}^{A_\ion,(m+1)} = |K|\frac{n_{\ion,K}^{(m)}}{\Delta t}+n_\gas|K| k_K^{(m+1)} n_{\mathrm{e, loc},K}^{(m)},
						\end{equation}
						\begin{equation}\label{eq:full_scheme_elec}
							|K|\frac{n_{\elec,K}^{(m+1)}}{\Delta t} +\sum_{\sigma\in\edges_K} F_{K,\sigma}^{D_\elec,(m+1)} + \sum_{\sigma\in\edges_K} F_{K,\sigma}^{A_\elec,(m+1)} = |K|\frac{n_{\elec,K}^{(m)}}{\Delta t}+n_\gas|K| k_K^{(m+1)} n_{\mathrm{e, loc},K}^{(m)},
						\end{equation}
						where $k_K^{(m+1)}$ is the average ionisation rate in cell $K$ at time $t^{(m+1)}$, and $n_{\mathrm{e, loc},K}^{(m)}$ is the value of $n_{\mathrm{e, loc}}^{(m)}$ computed at cell $K$. 
						
						Starting with $m\geq 1$, we then use the second-order approximation for the charge density in solving the Poisson equation. That is, instead of using \eqref{eq:full_scheme_Poisson}, we use a semi-implicit discretisation based on \eqref{eq:disc_Poisson_n1_semi_imp}, given by 
						\begin{equation}\label{eq:full_scheme_Poisson_withVent}
							\begin{aligned}
								\sum_{\edge \in \edgescv} F_{K,\sigma}^{D_V,(m+1)} &+q\Delta t \sum_{\edge \in \edgescv} F_{K,\sigma}^{C_V,(m+1)} \\
								&= q |K| (n_{\ion,K}^{(m)}-n_{\elec,K}^{(m)})+q\Delta t\bigg(\sum_{\sigma\in\edges_K}\big(F_{K,\sigma}^{D_\elec,(m)}-F_{K,\sigma}^{D_\ion,(m)})\bigg),
							\end{aligned}
						\end{equation}
						where 
						\begin{equation}\nonumber
							F_{K,\sigma}^{C_V,(m+1)} \approx \int_\sigma\Flux_\mathrm{c}^{(m+1)}\cdot \mathbf{n}_{K,\sigma} \mathrm{d}s
						\end{equation}
						is a discretisation of the semi-implicit correction flux $\Flux_\mathrm{c}$ defined in \eqref{eq:corr_flux}.
						
						We note here that we cannot immediately use \eqref{eq:full_scheme_Poisson_withVent} at the first time step due to the yet unknown mobility coefficients, correction and diffusive fluxes at time $t=0$. One natural option would be to use \eqref{eq:full_scheme_Poisson} for solving the Poisson equation at the first time step when $m=0$. Alternatively, we may opt to use a predictor-corrector method, e.g., starting with \eqref{eq:full_scheme_Poisson}, we get an initial estimate of the electric field, which will then be used for an initial estimate of the diffusion and mobility coefficients. These coefficients can then be used for computing the diffusive and correction fluxes, which in turn allows us to use \eqref{eq:full_scheme_Poisson_withVent} for solving the Poisson equation at time $t=0$. However, we notice in our numerical tests that this does not have a significant impact on the quality of the numerical solutions. Hence, for the simulations, we start with \eqref{eq:full_scheme_Poisson}.
					\end{subequations}
					
					\subsection{Boundary conditions and conservation of fluxes}
					
					We now see that there are $N_K$ equations (corresponding to the number of cells) involved in each of \eqref{eq:full_scheme_Poisson}-\eqref{eq:full_scheme_Poisson_withVent}. However, we have $N_K+N_\sigma$ unknowns for each of these equations: $N_K$ corresponding to the number of cells, and $N_\sigma$ corresponding to the number of edges. This is standard for hybrid finite volume methods, and the equations for the $N_\sigma$ unknowns along the edges are obtained by first imposing conservation of fluxes for interior edges. That is, if $\sigma$ is an edge shared by two cells $K$ and $L$, we should have that the total (diffusive and advective) flux $F_{K,\sigma}$ going from $K$ to $L$ via $\sigma$ is equal to the negative total flux $F_{L,\sigma}$ going from $L$ to $K$ via $\sigma$, i.e.
					\begin{equation}\label{eq:flux_cons}
						(F_{K,\sigma}^D+F_{K,\sigma}^A) + (F_{L,\sigma}^D+F_{L,\sigma}^A)=0.
					\end{equation}
					
					Finally, we impose boundary conditions on the remaining edges, which are located along the boundary of the domain. The Dirichlet and Neumann boundary conditions are straightforward to impose, so we only describe here how to impose the flux boundary conditions \eqref{eq:fluxBcI} and \eqref{eq:fluxBcE}. This is done by setting, for the corresponding boundary edges $\sigma\in\edges_K$,
					\begin{subequations}
						\begin{equation}\label{eq:fluxBCi}
							F_{K,\sigma}^{D_\ion,(m+1)} + F_{K,\sigma}^{A_\ion,(m+1)} =|\sigma|\bigg( (\mu_\ion^{(m+1)} \etensor^{(m+1)} \cdot  \mathbf{n}_{K,\sigma} )^+ n_{\ion,\sigma}^{(m+1)} + \frac{1}{4}v_{\mathrm{th, i}}n_{\ion,\sigma}^{(m+1)}\bigg).
						\end{equation} 
						\begin{equation}\label{eq:fluxBCe}
							\begin{aligned}
								F_{K,\sigma}^{D_\elec,(m+1)} + &F_{K,\sigma}^{A_\elec,(m+1)} \\&= |\sigma|\bigg( (\mu_\elec^{(m+1)}\mtensor_\elec^{(m+1)} \etensor^{(m+1)} \cdot  \mathbf{n}_{K,\sigma} )^+ n_{\elec,\sigma}^{(m+1)} + \frac{1}{4}v_{\mathrm{th, e}}n_{\elec,\sigma}^{(m+1)} \bigg)\\
								&\qquad-\gamma_\ion(F_{K,\sigma}^{D_\ion,(m+1)} + F_{K,\sigma}^{A_\ion,(m+1)}).
							\end{aligned}
						\end{equation}
					\end{subequations}
					Here, $n_{\elec,\sigma}$ and $n_{\ion,\sigma}$ are the values (to be determined) of $n_\elec$ and $n_\ion$ at $\sigma$, respectively. To complete the definition of the numerical scheme, we are then left to describe the choice of how to compute the discrete diffusive and advective fluxes.

					\subsection{Diffusive fluxes} \label{sec:diff_flux}
\begin{figure}[h]
						\begin{tabular}{cc}
							\includegraphics[width=0.325\linewidth]{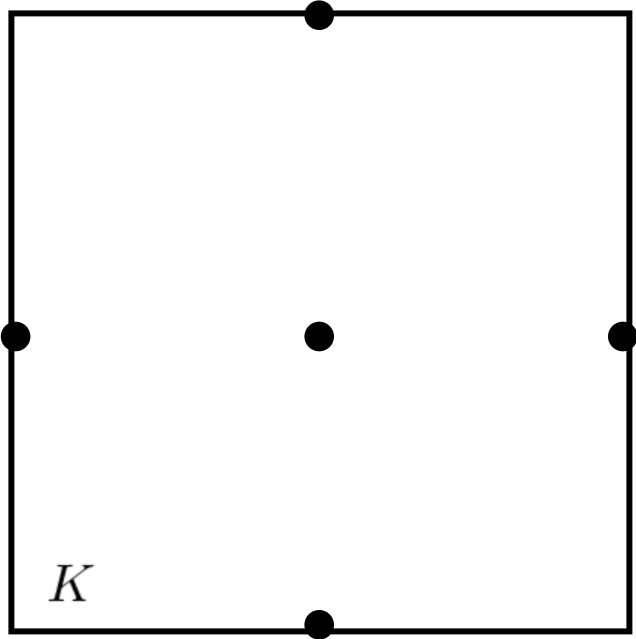} &\qquad
							\includegraphics[width=0.45\linewidth]{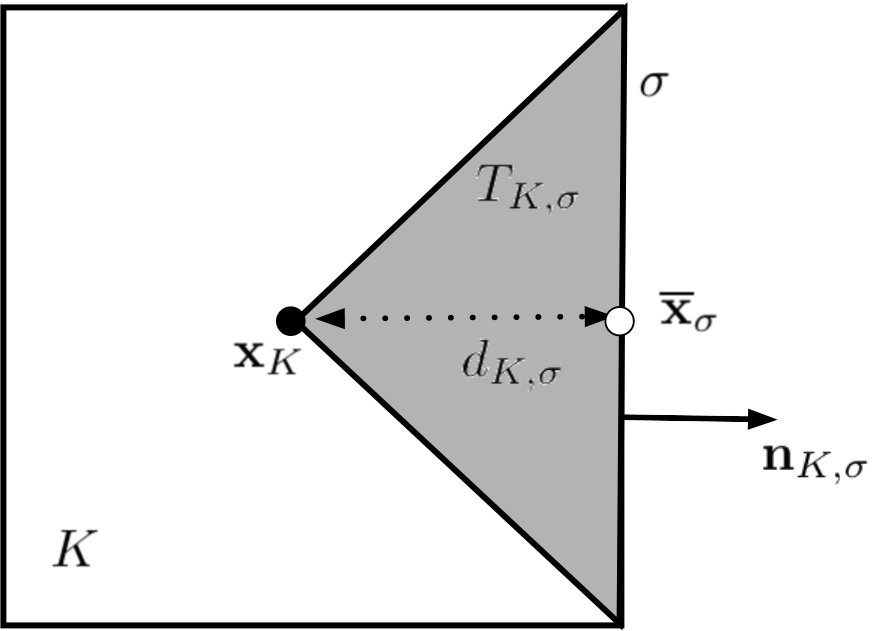}
						\end{tabular}
						\caption{Notation for a Cartesian cell in dimension $d=2$.} \label{fig.Cart_not}
					\end{figure}

					In this section, we discuss the discretisation of the diffusive fluxes $F_{K,\sigma}^D \approx -\int_\sigma \Lam \nabla c \cdot \mathbf{n}_{K,\sigma} \mathrm{d}s$. Here, we choose to use the hybrid mimetic mixed (HMM) method, due to its ability to handle anisotropic diffusion tensors on generic meshes, see e.g. \cite{dro-10-uni,EGH10-SUSHI}. Denoting by $v\in X_\disc$ a collection of discrete values, one on each cell, and one on each face (see Figure \ref{fig.Cart_not}, left), we then define the diffusive fluxes by the relation:
					\begin{equation}\label{eq:diff_flux}
						\begin{aligned}
							&\forall K\in\Omega_h\,,\;\forall v \in X_{\disc}\,,\\
							&\sum_{\edge \in \edgescv} F_{K,\sigma}^D (v_{K}-v_{\sigma}) = \int_{K} \Lam_K \nabla_{\disc} c(\mathbf{x}) \cdot \nabla_{\disc} v(\mathbf{x}) \mathrm{d}A ,
						\end{aligned}
					\end{equation}
					where $\nabla_{\disc}$ is a discrete gradient that is defined as follows: For $K\in\Omega_h$, we set
					\begin{subequations}
						\begin{align}
							\forall w\in X_{\disc}\,,\;\forall \x\in K\,,\nonumber&\\
							\grad_{\disc} w(\mathbf{x}) &= \ograd_{\cv} w + S_{K} w(\x),\label{def.grad}
						\end{align}
						
						\begin{equation}\label{def.grad_cons}
							\ograd_{\cv}w=\dfrac{1}{|\cv|}\sum_{\edge\in\edgescv} |\sigma|(w_{\edge}-w_K)\ncvedge.
						\end{equation}
						We see in \eqref{def.grad} that the discrete gradient consists of two components. The first component, $\ograd_{\cv} w$ in \eqref{def.grad_cons}, is a linearly exact reconstruction of the gradient. The second component, $S_{K}w(\x)$, is a stabilisation term, which is piecewise defined such that for each convex hull $(T_{K,\edge})_{\edge\in\edgescv}$ of $\edge$ and $\x_K$ (see Figure \ref{fig.Cart_not}, right), $S_{K}w(\x) = S_{K,\sigma} w$, with
						\begin{equation}\label{def.stab}
							S_{K,\sigma}w = \dfrac{\sqrt{2}}{\dcvedge}[w_{\edge}-w_{\cv}-\ograd_{\cv}w \cdot(\centeredge-\centercv)] \ncvedge.
						\end{equation}
					\end{subequations}
					We now illustrate the HMM method for Cartesian meshes. We note that even though the HMM is only illustrated for Cartesian meshes, it is also applicable to generic types of meshes, which will be demonstrated in the  tests in Section \ref{sec:Numtests_2D}. Since we work on square cells, the expressions in \eqref{def.grad_cons} and \eqref{def.stab} can be simplified. Firstly, we have for every edge $\sigma\in\edges_K$, $|\sigma|=h$. By adopting the compass notation and denoting by $\sigma_N$, $\sigma_S$, $\sigma_E$, $\sigma_W$ the north, south, east and west edges of cell $K$, respectively, we find that 
					\begin{equation}\nonumber
						\ograd_{\cv} w = \frac{1}{h}\begin{bmatrix}
							w_{\sigma_E}-w_{\sigma_W} \\
							w_{\sigma_N}-w_{\sigma_S}
						\end{bmatrix}.
					\end{equation}
					Moreover, we obtain the following simplified expression for the stabilisation term.
					\begin{equation}\nonumber
						S_{K,\sigma}w = \begin{cases}\frac{\sqrt{2}}{h} \big(w_{\sigma_E}-2w_K+w_{\sigma_W}\big) \ncvedge \quad \mathrm{for}\, \sigma = \sigma_E \, \mathrm{or} \, \sigma = \sigma_W, \\
							\frac{\sqrt{2}}{h}\big(w_{\sigma_N}-2w_K+w_{\sigma_S}\big) \ncvedge \quad \mathrm{for}\, \sigma = \sigma_N \, \mathrm{or} \, \sigma = \sigma_S.
						\end{cases}
					\end{equation}
					
					Finally, owing to \eqref{eq:diff_flux}, the diffusive flux $F_{K,\sigma}^D$ along the edge $\sigma$ of cell $K$ can be uniquely determined by choosing $v\in X_\disc$ such that $v_K=0, v_\sigma = -1$ and $v_{\sigma'}=0$ for $\sigma'\neq \sigma$. 
					
					\begin{example}[Computing the diffusive flux along an edge]As an example,  to compute the diffusive flux along the eastern edge $\sigma_E$ of a cell $K$, we choose $v\in X_\disc$ such that $v_K=0$, $v_{\sigma_E}=-1$, $v_{\sigma_N}=v_{\sigma_S}=v_{\sigma_W}=0$. Substituting into \eqref{eq:diff_flux} and denoting by $\Lam_K$ the value of the diffusion tensor at cell $K$, we then have 
						\begin{equation}\nonumber
							F_{K,\sigma_E}^D =\int_K \Lam_K \nabla_{\disc}c \cdot \nabla_{\disc} v \mathrm{d}A =\sum_{\sigma\in\edges_K} \int_{T_{K,\sigma}} \Lam_K \nabla_{\disc}c \cdot \nabla_{\disc} v \mathrm{d}A,
						\end{equation}
						with \begin{equation} \nonumber
							\Lam_K = \begin{bmatrix}
								\lambda_{K,11} & \lambda_{K,12}\\
								\lambda_{K,21} & \lambda_{K,22}
							\end{bmatrix}.
						\end{equation}
						We then compute 
						\begin{equation}\nonumber
							\nabla_{\disc} v  = \begin{cases}
								-\frac{1}{h} \mathbf{e}_x& \quad \mathrm{for} \, \sigma = \sigma_N \, \mathrm{or} \, \sigma = \sigma_S,\\
								\big(-\frac{1}{h}-\frac{\sqrt{2}}{h}\big)\mathbf{e}_x & \quad \mathrm{for} \, \sigma = \sigma_E, \\
								\big(-\frac{1}{h}+\frac{\sqrt{2}}{h}\big)\mathbf{e}_x & \quad \mathrm{for} \, \sigma = \sigma_W.
							\end{cases}
						\end{equation}
						Using definition \eqref{def.grad} of the discrete gradient and denoting the values of $c$ at the cell $K$ and its edges by $c_K, c_{\sigma_N}, c_{\sigma_S}, c_{\sigma_E}, c_{\sigma_W}$, respectively, we then compute the expression corresponding to $\nabla_{\disc}c$. Using the appropriate values of $\nabla_{\disc}c$ and $\nabla_{\disc} v$ for each of the regions $T_{K,\sigma}$, we obtain 
						\begin{equation}\nonumber
							F_{K,\sigma_E}^D = -1\bigg(2\lambda_{K,11}(c_{\sigma_E}-c_K)+\lambda_{K,12}(c_{\sigma_N}-c_{\sigma_S})\bigg).
						\end{equation}
						Here, we see that $	F_{K,\sigma_E}^D $ is an approximation of
						\begin{equation}\nonumber
							-\int_{\sigma_E} \Lam \nabla c \cdot \mathbf{n}_{K,\sigma_E} \mathrm{d}s = -\int_{\sigma_E} \big(\lambda_{K,11} c_x + \lambda_{K,12} c_y\big) \mathrm{d}s
						\end{equation} in the sense that $c_x$, $c_y$ are approximated by $\frac{2}{h}(c_{\sigma_E}-c_K)$ and  $\frac{1}{h}(c_{\sigma_N}-c_{\sigma_S})$, respectively. \qed
					\end{example}

					\subsection{Advective fluxes}
					We now discuss the discretisation of the advective fluxes $F_{K,\sigma}^A\approx\int_\sigma c\V \cdot \mathbf{n}_{K,\sigma} \mathrm{d}s$. The definition of the discrete advective fluxes is motivated by the Scharfetter-Gummel or exponential scheme \cite{SG69,AB11-FVCF}. Here, we briefly recall the Scharfetter-Gummel flux, and for simplicity of exposition, we consider an isotropic diffusion tensor $\Lam = \varepsilon \mathbf{I}$ and velocity field $\V=[v_1, 0]^T$. Defining the P\'eclet number 
					\begin{equation}\label{eq:Peclet1D}
						P = \frac{h v_1}{\varepsilon},
					\end{equation}
					the Scharfetter-Gummel flux across the eastern edge $\sigma_E$ of cell $K$ shared with cell $E$ is given by
					\begin{equation}\label{eq:SG_flux}
						F_{K,\sigma} = -\frac{\varepsilon|\sigma_E|}{h}\bigg(B(P)c_E-B(-P)c_K\bigg),
					\end{equation}
					where $B$ is the Bernoulli function with
					\begin{equation}\nonumber
						B(z)=\frac{z}{e^z-1}.
					\end{equation}
					We now note that the definition of the Scharfetter-Gummel flux \eqref{eq:SG_flux} is based on the P\'eclet number $P$. Hence, for a good definition of advective fluxes $F_{K,\sigma}^A$ to be used in \eqref{eq:full_scheme_ion} and \eqref{eq:full_scheme_elec}, we need a proper definition of a grid-based P\'eclet number for anisotropic advection-diffusion problems. In particular, for each $K\in\Omega_h, \edge\in\edgescv$, we define as in \cite[Section 3.2.2]{CT20-complete_flux} a local grid-oriented P\'eclet number
					\begin{equation}\label{eq:Peclet}
						P_{K,\sigma} = \frac{h V_{K,\sigma}}{2\lambda_{K,\sigma}},
					\end{equation} 
					with 
					\begin{equation}\nonumber
						\begin{aligned}
							V_{K,\sigma} &= \frac{1}{|\sigma|}\int_\sigma \V \cdot \mathbf{n}_{K,\sigma} \mathrm{d}s,\\
							\lambda_{K,\sigma} &= \mathbf{n}_{K,\sigma}^T \Lam_K \mathbf{n}_{K,\sigma}.
						\end{aligned}
					\end{equation}
					Here, the quantities $V_{K,\sigma}$ and $\lambda_{K,\sigma}$ measure the strength of advection and diffusion across the outward normal $\mathbf{n}_{K,\sigma}$, respectively; hence allowing the P\'eclet number $P_{K,\sigma}$ to capture the relative strength of advection over diffusion along $\mathbf{n}_{K,\sigma}$ properly. As a remark, we note that for isotropic diffusion, the expression \eqref{eq:Peclet} of the P\'eclet number boils down to the classical definition \eqref{eq:Peclet1D} of the P\'eclet number.
					
					Since the diffusive fluxes have already been defined in Section \ref{sec:diff_flux}, we consider only the advective component of the Scharfetter-Gummel flux. This is done by following the ideas in \cite{AB85-MFEM,VDM11-adv-diff,CD09-FV,CHLM22-longtermFV}, where instead of the Bernoulli function, we extract the advective component of the Scharfetter-Gummel fluxes by defining 
					\begin{equation}\nonumber
						A_{\mathrm{sg}}(z) = B(-z)-1.
					\end{equation} The advective fluxes are then given by  
					\begin{equation}\label{eq:advFlux}
						\begin{aligned}
							F_{K,\sigma}^A &=  \frac{2\lambda_{K,\sigma}|\sigma|}{h}\bigg(A_{\mathrm{sg}}\big(P_{K,\sigma}\big) c_K - A_{\mathrm{sg}}\big(-P_{K,\sigma}\big) c_\sigma\bigg).
						\end{aligned}
					\end{equation}
					
					As a remark, the accuracy and convergence of numerical solutions from the combination of the HMM with the Scharfetter-Gummel scheme for advection-diffusion equations have already been studied in \cite{VDM11-adv-diff}.
					
					\begin{remark}[Charge conservation]
						Looking at \eqref{eq:full_scheme_Poisson_withVent}, we see that due to \eqref{eq:drho_dt}, the integral of the term $\frac{\partial \rho}{\partial t}$ over a cell can actually be expressed in terms of the integral of the charge density fluxes over the cell's faces. This suggests that our usage of a finite volume discretisation in space ensures that a positive contribution of a given cell is balanced by a negative contribution from its neighboring cell, i.e. the charge is always conserved for internal faces. We note however that before reaching steady state, charge is not necessarily conserved at the boundary, due to the fact that Dirichlet boundary conditions \eqref{eq:DBc} are imposed for the potential. This means that a net charge is present when integrating the charge density over the control volume at the boundary.
					\end{remark}

					\section{Summary and implementation of numerical scheme}\label{sec:Imp}
					
					\noindent In this section, we present a summary of the numerical scheme for the magnetised transport problem described by the model system of equations \eqref{eq:Poisson}, \eqref{eq:magnetised_transport_ion}, and \eqref{eq:magnetised_transport_electron}. Starting with $m=0$ and initial ion and electron density profiles of $n_\ion^{(0)}$ and $n_\elec^{(0)}$, respectively, the idea is to sequentially solve, the potential $V^{(m+1)}$, followed by the ion density $n_{\ion}^{(m+1)}$ and then the electron density $n_\elec^{(m+1)}$. This process is repeated until we reach a time $T=t^{(m+1)}$ such that the solution is already at steady state. Numerically, the solution is said to be at steady state whenever 
					\begin{equation}\label{req:steady_state}
						\frac{\norm{V^{(m+1)}-V^{(m)}}{}}{\Delta t}<\mathrm{tol}, \quad \frac{\norm{n_\ion^{(m+1)}-n_\ion^{(m)}}{}}{\Delta t}<\mathrm{tol}, \quad \frac{\norm{n_\elec^{(m+1)}-n_\elec^{(m)}}{}}{\Delta t}<\mathrm{tol}.
					\end{equation}
					We now elaborate the details of the computations involved in the first few steps of the scheme. 
					We start by computing the potential $V^{(1)}$ via solving \eqref{eq:full_scheme_Poisson} with diffusive fluxes \eqref{not:diff_fluxV} defined as in  \eqref{eq:diff_flux}. Under the assumption that the initial ion and electron densities are equal, we then solve at time $m=0$, for each $K\in\Omega_h$, the balance of fluxes
					\begin{subequations}
						\begin{equation} \label{eq:pot_t1}
							-2\varepsilon(V^{(1)}_{\sigma_W}-2V^{(1)}_K+V^{(1)}_{\sigma_E}) -2\varepsilon(V^{(1)}_{\sigma_S}-2V^{(1)}_K+V^{(1)}_{\sigma_N})=0.
						\end{equation}
						This is solved together with the conservation of fluxes  for interior edges $\sigma$ between cells $K$ and $L$, given by
						\begin{equation}\label{eq:pot_fluxcons}
							V^{(1)}_\sigma = \frac{1}{2}(V^{(1)}_K+V^{(1)}_L).
						\end{equation}
						Finally, to complete the set of equations for the potential $V^{(1)}$, we impose, for the appropriate edges, Dirichlet boundary conditions,
						\begin{equation}\label{eq:pot_DBC}
							V^{(1)}_\sigma = V_\mathrm{a} \quad \mathrm{ or } \quad V^{(1)}_\sigma = V_\mathrm{c},
						\end{equation}
						and Neumann boundary conditions, given by
						\begin{equation}\label{eq:pot_NBC}
							V^{(1)}_\sigma = V^{(1)}_K.
						\end{equation}
					\end{subequations}
					The linear system \eqref{eq:pot_t1}--\eqref{eq:pot_NBC} then fully defines $V^{(1)}$. As a remark, we note that due to the absence of anisotropy in the Poisson equation, the HMM method on Cartesian meshes boils down to a standard central difference scheme.  We also note that using only the consistent part \eqref{def.grad_cons} of the discrete gradient will yield an electric field $\etensor^{(1)}$ that does not satisfy the homogeneous Neumann boundary condition \eqref{eq:NBc}. Hence, we reconstruct the HMM discrete gradient as in \eqref{def.grad} in order to obtain the value of the electric field $\etensor^{(1)}$ on each cell $K\in\Omega_h$.
					
					At first glance, it may seem that the hybrid scheme is very expensive to implement, since it involves solving a system of $N_K+N_\sigma$ equations in $N_K +N_\sigma$ unknowns. However, by looking at the first $N_K$ equations \eqref{eq:pot_t1} corresponding to the cells $K\in\Omega_h$, we see that they are fully local in the sense that the value of $V$ on a cell $K$ only depends upon the values of $V$ on its edges. Hence, static condensation can be applied for an efficient implementation of the scheme, i.e., in terms of the matrix system 
					\begin{equation}\nonumber
						A\mathbf{x}=\mathbf{b}
					\end{equation}
					that needs to be solved, we actually realise that upon partitioning and writing 
					\begin{equation}\nonumber
						A=\begin{bmatrix}
							A_{11} & A_{12} \\
							A_{21} & A_{22}
						\end{bmatrix},
					\end{equation}
					where $A_{11}$ and $A_{12}$ contains the first $N_K$ equations \eqref{eq:pot_t1} for  the unknowns $V_K$, we see that $A_{11}$ is actually diagonal, and hence, the main cost of the scheme only involves solving a system of $N_\sigma$ equations for $N_\sigma$ unknowns $V_\sigma$.
					
					Using the value of $\etensor^{(m+1)}$, the ionisation rates $k$ and the mobility coefficients $\mu_\elec$ and $\mu_\ion$ are computed by the lookup tables as in Figures \ref{fig.lookup_k} and \ref{fig.lookup_mu}. The diffusion coefficients $D_\ion$ and $D_\elec$ are then calculated according to the definitions \eqref{eq:Di_ion} and \eqref{eq:Di_elec}, respectively. Following this, we compute the local electron density on each cell $K\in\Omega_h$ as in \eqref{eq:disc_ne_loc}. These quantities allow us to compute the diffusive and advective fluxes \eqref{not:diff_fluxi}-\eqref{not:adv_fluxe} by using the formulations \eqref{eq:diff_flux} and \eqref{eq:advFlux}. We then substitute the appropriate expressions corresponding to \eqref{not:diff_fluxi} and \eqref{not:adv_fluxi} into \eqref{eq:full_scheme_ion} to find $n_{\ion}^{(m+1)}$. The system of equations \eqref{eq:full_scheme_ion} used to compute $n_\ion$ are given by, for each $K\in\Omega_h$,
					\begin{flalign}
						|K|\frac{n_{\ion,K}^{(m+1)}}{\Delta t}
						&-2D_\ion\bigg(n_{\ion,\sigma_W}^{(m+1)}-2n_{\ion,K}^{(m+1)}+n_{\ion,\sigma_E}^{(m+1)}\bigg)&& \nonumber \\
						&-2D_\ion\bigg(n_{\ion,\sigma_S}^{(m+1)}-2n_{\ion,K}^{(m+1)}+n_{\ion,\sigma_N}^{(m+1)}\bigg) && \nonumber\\
						&+ 2D_\ion\bigg(\big(A_{\mathrm{sg}}(P_{K,\sigma_E})+A_{\mathrm{sg}}(-P_{K,\sigma_E})\big)n_{\ion,K}^{(m+1)}\bigg)&& \nonumber\\
						&+2D_\ion\bigg(\big(A_{\mathrm{sg}}(P_{K,\sigma_N})+A_{\mathrm{sg}}(-P_{K,\sigma_N})\big)n_{\ion,K}^{(m+1)}\bigg)&& \nonumber\\
						&-2D_\ion\bigg(A_{\mathrm{sg}}(P_{K,\sigma_E})n_{\ion,\sigma_W}^{(m+1)}+A_{\mathrm{sg}}(-P_{K,\sigma_E})n_{\ion,\sigma_E}^{(m+1)}\bigg)&& \nonumber\\
						&-2D_\ion\bigg(A_{\mathrm{sg}}(P_{K,\sigma_N})n_{\ion,\sigma_S}^{(m+1)}+A_{\mathrm{sg}}(-P_{K,\sigma_N})n_{\ion,\sigma_N}^{(m+1)}\bigg)
						&& \nonumber\\
						&\,\,\,= |K|\frac{n_{\ion,K}^{(m)}}{\Delta t}+|K|k_K^{(m+1)} n_{\elec,\mathrm{loc},K}^{(m)} n_\gas, \nonumber
					\end{flalign}
					where, denoting by $E_x$ and $E_y$ the $x$- and $y$- components of the electric field, respectively, the P\'eclet numbers are given by $P_{K,\sigma_E}= h\mu_\ion E_x^{(m+1)}/(2D_\ion), P_{K,\sigma_N}= h\mu_\ion E^{(m+1)}_y/(2D_\ion)$. Here, we also used the fact that $P_{K,\sigma_W}=-P_{K,\sigma_E}$, $P_{K,\sigma_S}=-P_{K,\sigma_N}$.
					Following this, the conservation of flux conditions are obtained by substituting, for interior edges, the appropriate expressions for \eqref{not:diff_fluxi} and \eqref{not:adv_fluxi} in \eqref{eq:flux_cons}. For the boundary edges corresponding to electrodes, the flux boundary conditions \eqref{eq:fluxBCi} are applied. Finally, the homogeneous Neumann boundary conditions require that
					\begin{equation}\nonumber
						n_{\ion,\sigma}^{(m+1)} = n_{\ion,K}^{(m+1)}.
					\end{equation}
					We then use $n_\ion^{(m+1)}$ to compute the secondary electron emission $\Flux_\ion^{(m+1)} \cdot \mathbf{n}$ for \eqref{eq:fluxBCe}.  Lastly, a similar process may be used to assemble the system of equations needed to compute $n_\elec^{(m+1)}$, i.e., by assembling the $N_K$ equations \eqref{eq:full_scheme_elec}, and imposing the appropriate flux conservation conditions \eqref{eq:flux_cons} and boundary conditions. The main difference between the equations for $n_\elec$ and $n_\ion$ comes from the presence of the magnetic tensor $\mtensor_\elec$ in \eqref{eq:full_scheme_elec}. Due to anisotropy, the expression for the diffusive flux of $n_\elec$ is different from the one obtained for $n_\ion$.  Moreover, this requires us to use the P\'eclet number as defined in \eqref{eq:Peclet} for the advective fluxes of $n_\elec$.
					
					For $m\geq 1$, we use \eqref{eq:full_scheme_Poisson_withVent} instead of \eqref{eq:full_scheme_Poisson} to solve the Poisson equation for the potential $V$ and electric field $\etensor$, which will be used for computing the ion and electron densities in the next time steps. 
					
					\section{Numerical tests}\label{sec:Numtests}
					\noindent In this section, we perform numerical tests using the scheme \eqref{eq:full_scheme} described above for the model system of equations \eqref{eq:Poisson}, \eqref{eq:magnetised_transport_ion}, and \eqref{eq:magnetised_transport_electron}. Here, the domain involves 2 parallel plates that are 0.02m  apart, which is given by $\Omega = (0,0.02)\times (0,0.02)$. We start with a simple test case, for which there is an anode on the left wall, and a cathode on the right wall. In particular, we set 
					\begin{equation}\nonumber
						\begin{aligned}
							V = V_\mathrm{a} = -300 \mathrm{V} \quad &\mathrm{on} \quad x=0,\\
							V = V_\mathrm{c} = 0\mathrm{V} \quad &\mathrm{on} \quad x=0.02.
						\end{aligned}
					\end{equation}
					For all test cases, we use the following parameters in Table \ref{tab:test_param}. The thermal velocities of the ions and electrons presented below were calculated by the formula 
					\begin{equation}\nonumber
						v_{\mathrm{th, p}} = \sqrt{\frac{8k_\mathrm{B}T_\ptc}{\pi m_\ptc}}
					\end{equation} with $m_\ion = 6.6462 \times 10^{-27} kg$ and $m_\elec = 9.1093 \times 10^{-31}kg$.
					\begin{table}[h!]
						\caption{Parameters for the test.}\label{tab:test_param}
						\centering
						\begin{tabular}{|ccc|}
							\hline
							temperature of the background gas & $T_\gas$ & $300K$\\
							density of the background gas &    $n_\gas$ &  $1.9\times 10^{23} m^{-3}$   \\
							
							thermal velocity, electron     & $v_{\mathrm{th, e}}$ & $7.7277\times 10^5 m/s$ \\
							
							thermal velocity, ion     & $v_{\mathrm{th, i}}$ & $1.2598\times 10^3 m/s$ \\
							emission coefficient & $\gamma_\ion$ & 0.2 \\
							initial density, electron     & $n_\elec(\x,0)$ & $5\times 10^{13} m^{-3}$ \\
							
							initial density, ion    & $n_\ion(\x,0)$ & $5\times 10^{13} m^{-3}$\\
							\hline
						\end{tabular} 
					\end{table}
					
					\subsection{1D test}
\begin{figure}[h]
						\begin{tabular}{cc}
							\includegraphics[width=0.45\linewidth]{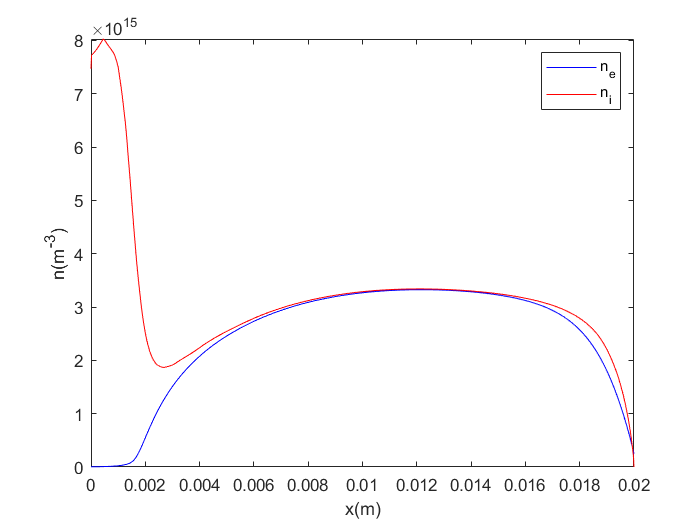} &
							\includegraphics[width=0.45\linewidth]{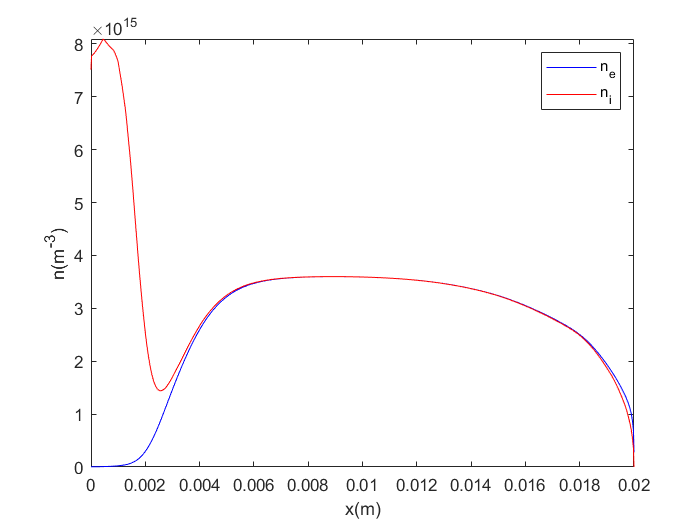}
						\end{tabular}
						\caption{Steady state density profiles, test case 1, mesh with 512 cells (Left: scheme \eqref{eq:full_scheme}, Right: Scharfetter-Gummel scheme).} \label{fig.1Dtest_fine}
					\end{figure}

					We start with a 1D test case, obtained by setting $\mtensor_\elec=\mathbf{I}$, and consider $\Omega= (0,0.02)$, partitioned uniformly. This will be referred to as test case 1. We compare in Figure \ref{fig.1Dtest_fine} the steady-state density profiles between our proposed scheme and the classical Scharfetter-Gummel scheme \cite{SG69} on a fine mesh with 512 cells, whilst taking a small time step of $\Delta t= 2$e-9s. Here, we set the tolerance value to indicate that the densities have reached steady state in \eqref{req:steady_state} to be $\mathrm{tol}=0.01$.  Typically, this is achieved at time $t=1$e-5s.
					
\begin{figure}[h]
						\begin{tabular}{cc}
							\includegraphics[width=0.45\linewidth]{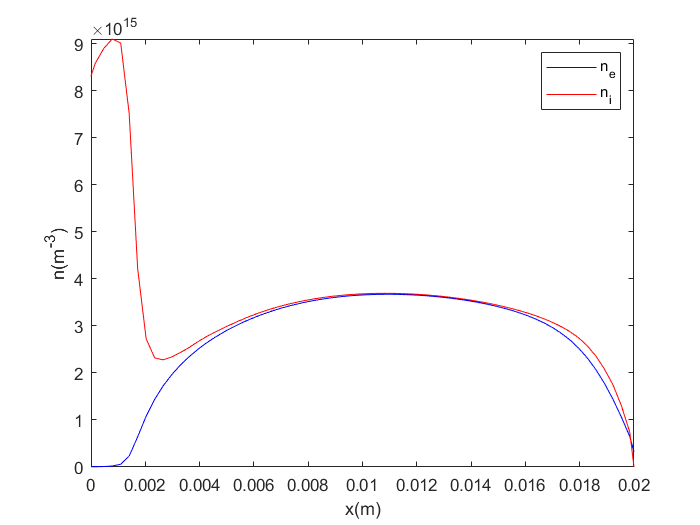} &
							\includegraphics[width=0.45\linewidth]{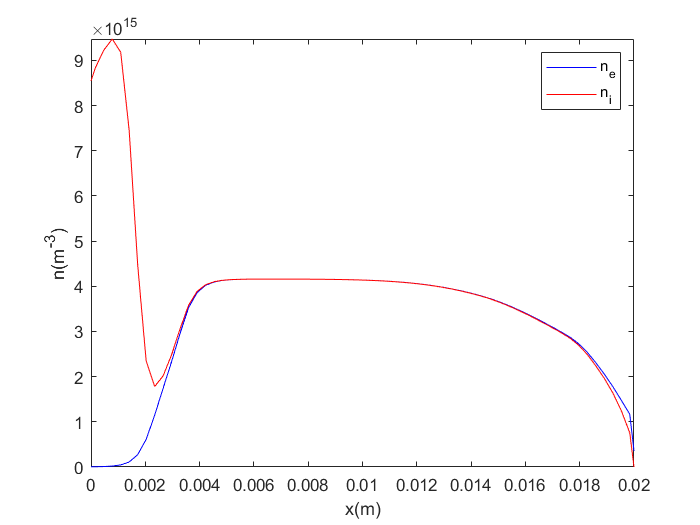}
						\end{tabular}
						\caption{Steady state density profiles, test case 1, mesh with 64 cells (Left: scheme \eqref{eq:full_scheme}, Right: Scharfetter-Gummel scheme).} \label{fig.1Dtest_coarse}
					\end{figure}

					In Figure \ref{fig.1Dtest_fine}, we see that the numerical solutions from both methods give a similar shape for the density profiles. To be specific, the ion densities peak on the interval $(0,0.002)$, and a quasi-neutral state is reached for $x>0.002$. However, upon having a closer look at Figure \ref{fig.1Dtest_fine}, right, we see that the numerical solution obtained from the Scharfetter-Gummel scheme exhibits a steeper dip in the ion density before reaching quasi-neutrality. Moreover, the ion and electron densities in this region are slightly larger than that of scheme \eqref{eq:full_scheme}. In particular, we have that the maximum ion and electron densities in this region are $3.6\times10^{15}m^{-3}$ for the Scharfetter-Gummel scheme, which is approximately 10\% larger than $3.2\times10^{15}m^{-3}$, obtained from scheme \eqref{eq:full_scheme}.  We now compare the solution profiles on a coarser mesh, now consisting of 64 cells in Figure \ref{fig.1Dtest_coarse}.

					In Figure \ref{fig.1Dtest_coarse}, we observe that the numerical solutions on the coarse mesh still exhibit the same behavior as the solutions on the fine mesh. Namely, the ion densities still peak in the interval $(0,0.002)$ and a quasi-neutral state is still observed for $x>0.002$. However, we see for both schemes an increase in the ion and electron densities compared to the solutions obtained on the fine mesh. This is not unexpected, since the solutions on a coarse mesh are in most cases less accurate than the ones on a fine mesh. We note however that the solution peaks only rise by around 12.5\%, which is an acceptable trade-off, since the computational costs are reduced by a factor of 8.
					
					Due to the use of a second-order approximation of the charge density in the Poisson equation by solving \eqref{eq:full_scheme_Poisson_withVent}, we may take a larger time step of $\Delta t = 4$e-8s for both methods, and still obtain results that are essentially the same as those presented in Figures \ref{fig.1Dtest_fine} and \ref{fig.1Dtest_coarse}.
					
					\subsection{2D tests} \label{sec:Numtests_2D}
\begin{figure}[h]
						\begin{tabular}{cc}
							\includegraphics[width=0.45\linewidth]{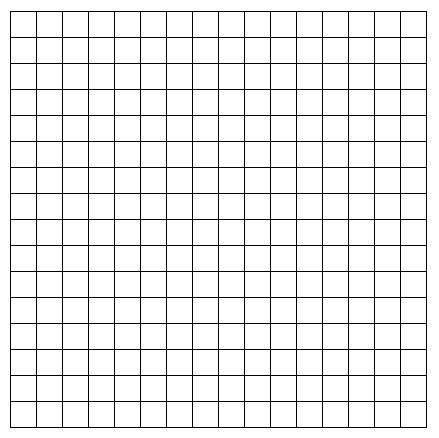} &
							\includegraphics[width=0.45\linewidth]{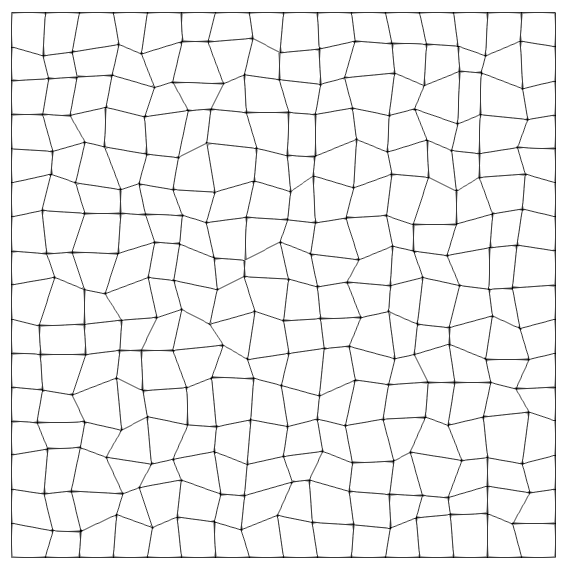}
						\end{tabular}
						\caption{Meshes in 2D (Left: Cartesian, Right: perturbed Cartesian)} \label{fig.2Dmeshes}
					\end{figure}

					We now proceed to a 2D test. For these tests, we will present our results on Cartesian meshes (Figure \ref{fig.2Dmeshes}, left). In order to illustrate that our scheme can also be applied to more generic types of meshes, we also ran tests on perturbed Cartesian meshes (Figure \ref{fig.2Dmeshes}, right), constructed following the guidelines provided in \cite{L10-monotoneFV}. That is, starting with a uniform Cartesian mesh (Figure \ref{fig.2Dmeshes}, left), if the maximum diameter of the cells are given by $h$, then the internal nodes $(x,y)$ are perturbed randomly by taking
						\[
						\hat{x} := x + 0.4\beta_x h, \quad \hat{y} := y +0.4\beta_y h,
						\]
						where $\beta_x,\beta_y$ are random values from a uniform distribution on $(-0.5,0.5)$.

					We then consider a two-dimensional magnetic field. That is, we take $\mathbf{B}$ as in \eqref{eq:mag_angle}, with 
					\begin{equation}\nonumber
						\norm{\mathbf{B}(\x)}{} = 0.2 \min\bigg(1,\frac{2x}{0.02}\bigg)\min\bigg(1,\frac{2y}{0.02}\bigg) .
					\end{equation}
					Physically, this means that the magnetic field grows stronger as we get closer to the upper and right parts of the domain, with a maximum strength of 0.2. We then perform numerical tests with magnetic field angles of $\theta=45^\circ$, and $\theta=90^\circ$ in \eqref{eq:mag_angle}, respectively on a uniform and perturbed grid consisting of $64\times 64$ cells, with a time step of $\Delta t = 4$e-8s. We will refer to this as test case 2. We will only present one figure for each of the tests, as the results obtained on the perturbed mesh are essentially the same as those obtained on the uniform mesh.
					\begin{figure}[h]
						\begin{tabular}{cc}
							\includegraphics[width=0.45\linewidth]{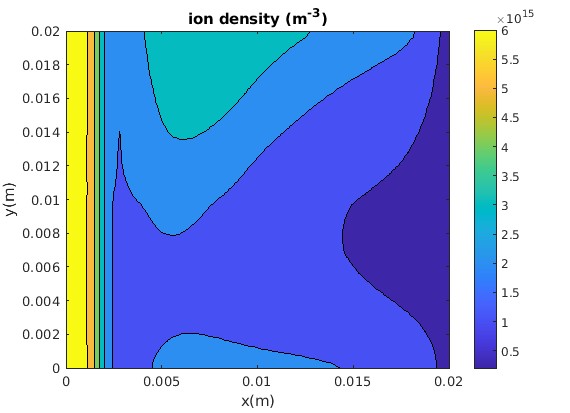} &
							\includegraphics[width=0.45\linewidth]{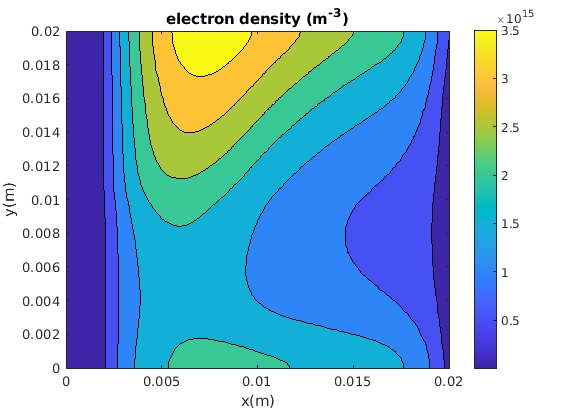}
						\end{tabular}
						\caption{Steady state density profiles, test case 2, magnetic field with $\theta =45^\circ$  (Left: $n_\ion$, Right: $n_\elec$).} \label{fig.2Dtest_45}
					\end{figure}
					
					\begin{figure}[h]
						\begin{tabular}{cc}
							\includegraphics[width=0.45\linewidth]{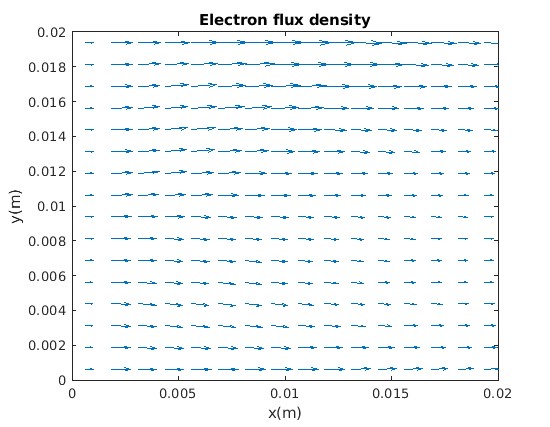} &
							\includegraphics[width=0.45\linewidth]{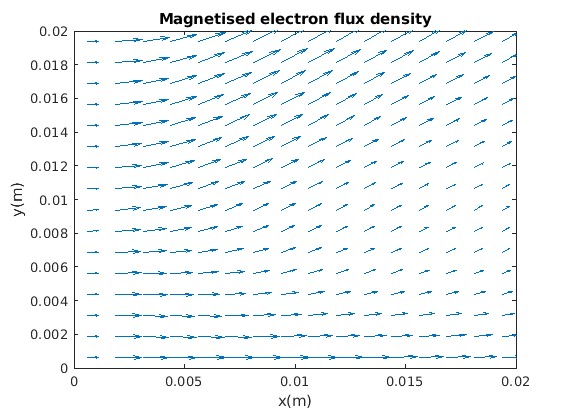}
						\end{tabular}
						\caption{Electron flux density, test case 2, magnetic field with $\theta = 45^\circ$ .} \label{fig.2Dtestfluxes_45}
					\end{figure}
					
					\begin{figure}[h]
						\begin{tabular}{cc}
							\includegraphics[width=0.45\linewidth]{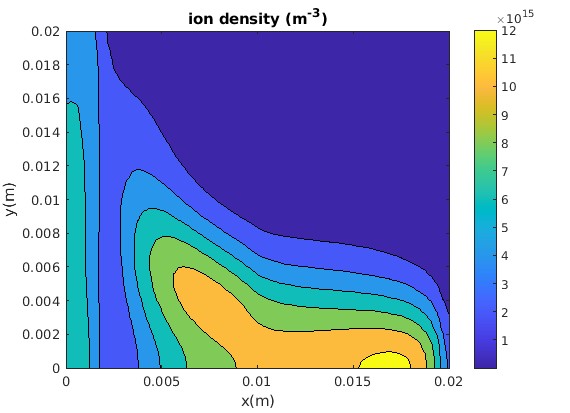} &
							\includegraphics[width=0.45\linewidth]{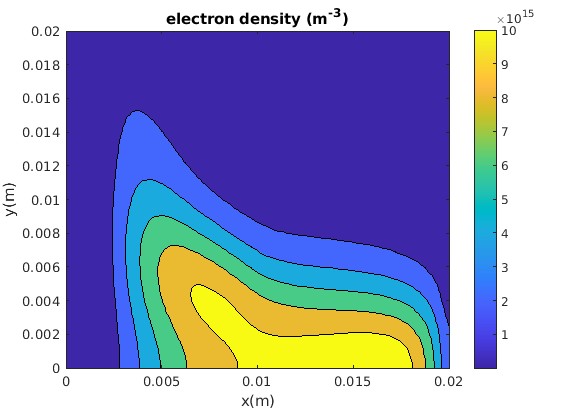}
						\end{tabular}
						\caption{Steady state density profiles, test case 2, magnetic field with $\theta = 90^\circ$ (Left: $n_\ion$, Right: $n_\elec$).} \label{fig.2Dtest_90}
					\end{figure}
					\begin{figure}[h]
						\begin{tabular}{cc}
							\includegraphics[width=0.45\linewidth]{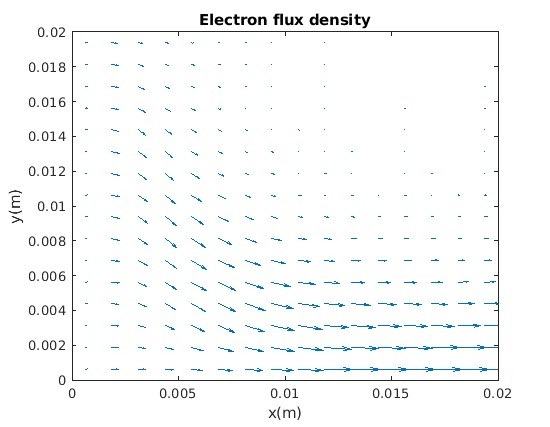} &
							\includegraphics[width=0.45\linewidth]{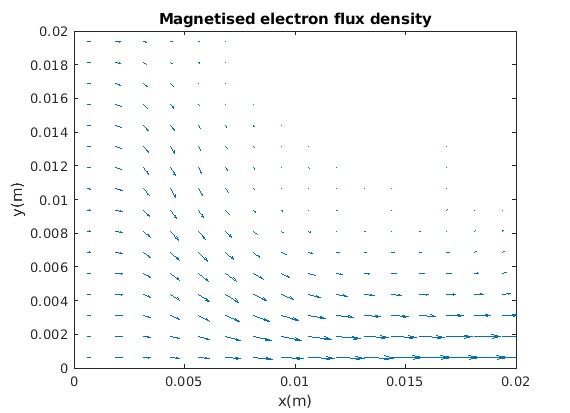}
						\end{tabular}
						\caption{Electron flux density, test case 2, magnetic field with $\theta =90^\circ$.} \label{fig.2Dtestfluxes_90}
					\end{figure}
					
\begin{figure}
						\begin{tabular}{cc}
							\includegraphics[width=0.45\linewidth]{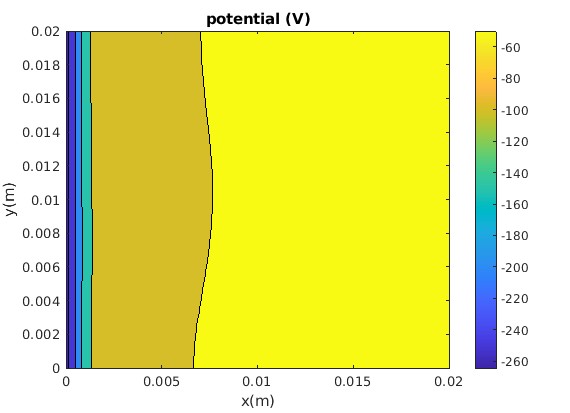} &
							\includegraphics[width=0.45\linewidth]{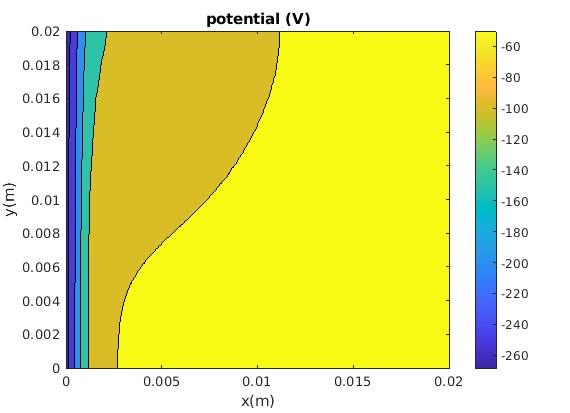}
						\end{tabular}
						\caption{Potential $V$, test case 2, magnetic field angle (Left: $\theta=45^\circ$, Right: $\theta=90^\circ$).} \label{fig.2Dtest_pot}
					\end{figure}

					In both Figures \ref{fig.2Dtest_45} and \ref{fig.2Dtest_90}, we see that the electron and ion densities are sparse on the upper right region of the domain. This is due to the fact that the electrons did not have sufficient time to go to the opposite electrode. Physically, this is expected, because the magnetic field creates a barrier with an angle of $45^\circ$ and $90^\circ$, respectively, in the upper right half of the domain. This barrier separates the electrons that have a smaller loss term at the upper boundary than at the electrode for a $45^\circ$ and $90^\circ$ angle. We now discuss the behaviour of the density profiles for the $45^\circ$ angle in more detail. In particular, we discuss separately the effects of the magnetic field at the upper right, central, and left regions, respectively (see Figure \ref{fig.2Dtest_45}, right). Firstly, the loss at the upper right corner of the domain is equal to the wall loss. Going towards the upper center, this wall loss at the upper boundary decreases because the electrons are not fully magnetised. Additionally, the influx from other parts of the domain is high at this location. Finally, at the upper left, the electron flux density from other parts of the domain is lower due to the smaller magnetisation and therefore the density decreases. Figures \ref{fig.2Dtestfluxes_45} and \ref{fig.2Dtestfluxes_90} show the effect of the magnetic field angle on the fluxes $\Flux_\elec$. In particular, we see in Figure \ref{fig.2Dtestfluxes_45} that after magnetisation, the fluxes are now oriented in an angle of $45^\circ$ at the upper right region of the domain. On the other hand, Figure \ref{fig.2Dtestfluxes_90} shows that by applying a $90^\circ$ magnetic field, the fluxes which were initially traveling from left to right horizontally has been decreased significantly. As discussed in \eqref{eq:mag_field_perp}, this is expected since the flux is orthogonal to the magnetic field. The discussion above and the results in Figures \ref{fig.2Dtest_45}--\ref{fig.2Dtestfluxes_90} show that the scheme \eqref{eq:full_scheme} can capture the shape of the solution profile and the physics properly on both uniform and distorted meshes. Of course, as seen in the 1D test, if we want a more accurate depiction of the peaks of the solution profile, we need to work on a finer mesh. We also present in Figure \ref{fig.2Dtest_pot} the potential $V$ in both tests. 
					
\begin{figure}
						\begin{tabular}{cc}
							\includegraphics[width=0.45\linewidth]{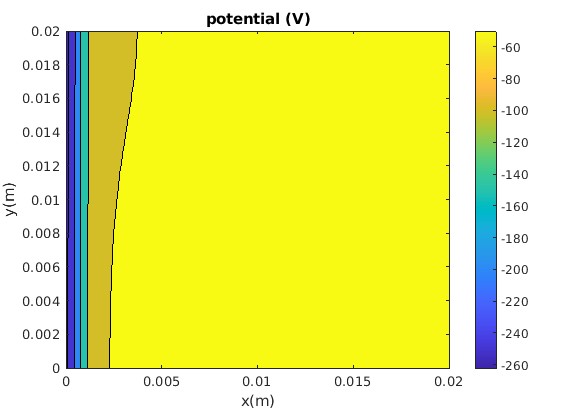} &
							\includegraphics[width=0.45\linewidth]{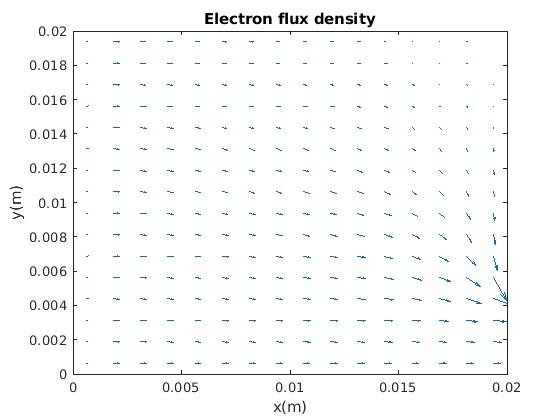}
						\end{tabular}
						\caption{Test case 3, no magnetic field (Left: potential $V$, Right: electron flux density $\Flux_\elec$).} \label{fig.2Dtest2fluxes_0}
					\end{figure}

\begin{figure}
						\begin{tabular}{cc}
							\includegraphics[width=0.45\linewidth]{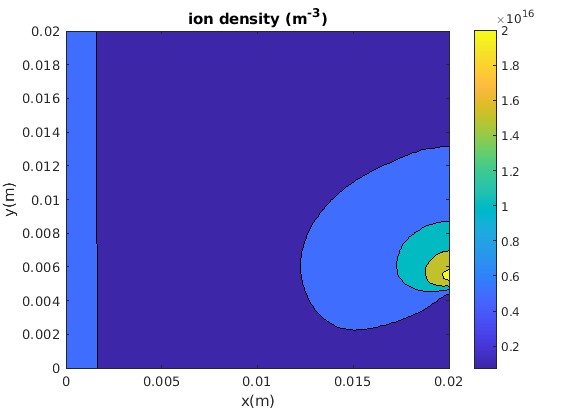} &
							\includegraphics[width=0.45\linewidth]{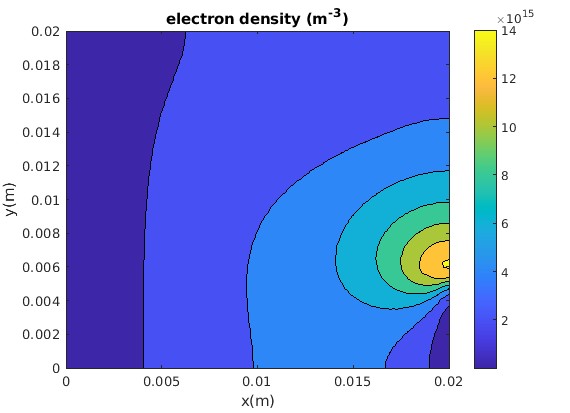}
						\end{tabular}
						\caption{Steady state density profiles, test case 3, no magnetic field (Left: $n_\ion$, Right: $n_\elec$).} \label{fig.2Dtest2_0}
					\end{figure}

					In order to further test the limits of the scheme, we consider a test case for which the boundary is not fully a cathode. To do so, we slightly alter the boundary conditions of the Poisson equation. In particular, instead of purely having a cathode and setting $V=0$V on $x=0.02$, we impose instead 
					\begin{equation}\nonumber
						\begin{aligned}
							V&=0 \quad \mathrm{on} \quad x=0.02, y\in\bigg(0,\frac{1}{4}\times 0.02\bigg)\\
							\nabla V \cdot \mathbf{n} &= 0 \quad \mathrm{on } \quad x=0.02, y\in\bigg(\frac{1}{4}\times 0.02,0.02\bigg).
						\end{aligned}
					\end{equation}
					Note that since the upper part of the right wall is no longer a cathode, the corresponding boundary conditions for the ion and electron continuity equations at this region should also be changed. That is, we set, for $\ptc = \ion, \elec$,
					\begin{equation}\nonumber
						\begin{aligned} 
							\nabla n_\ptc \cdot \mathbf{n} &= 0 \quad \mathrm{on } \quad x=0.02, y\in\bigg(\frac{1}{4}\times 0.02,0.02\bigg).\\
						\end{aligned}
					\end{equation}
					We will refer to this as test case 3. As with test case 2, we perform our numerical simulations on a grid with $64 \times 64$ cells, with a time step of $\Delta t = 4$e-8s. Note that upon changing the boundary conditions from test case 2, Figures \ref{fig.2Dtest2fluxes_0} and  \ref{fig.2Dtest2_0} exhibit two-dimensional potential and density profiles, even with no magnetisation (setting $\mtensor_\elec=\mathbf{I}$).
					
\begin{figure}
						\begin{tabular}{cc}
							\includegraphics[width=0.45\linewidth]{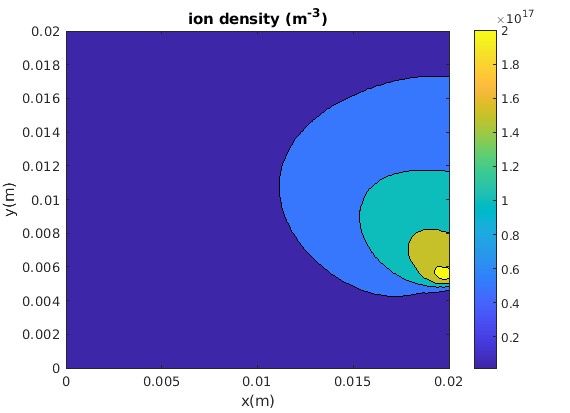} &
							\includegraphics[width=0.45\linewidth]{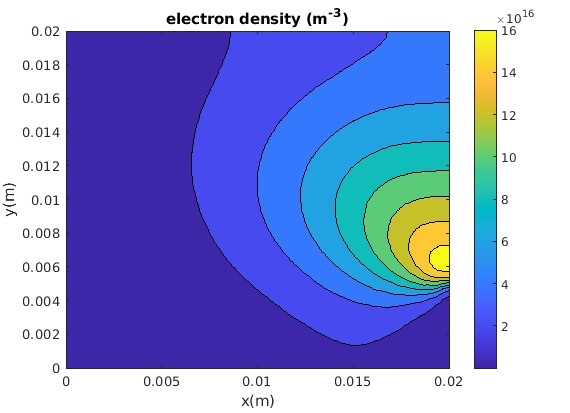}
						\end{tabular}
						\caption{Steady state density profiles, test case 3, $45^\circ$ magnetic field (Left: $n_\ion$, Right: $n_\elec$).} \label{fig.2Dtest2_45}
					\end{figure}

\begin{figure}
						\begin{tabular}{cc}
							\includegraphics[width=0.45\linewidth]{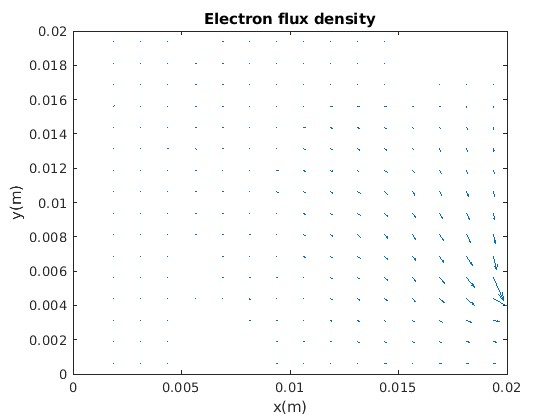} &
							\includegraphics[width=0.45\linewidth]{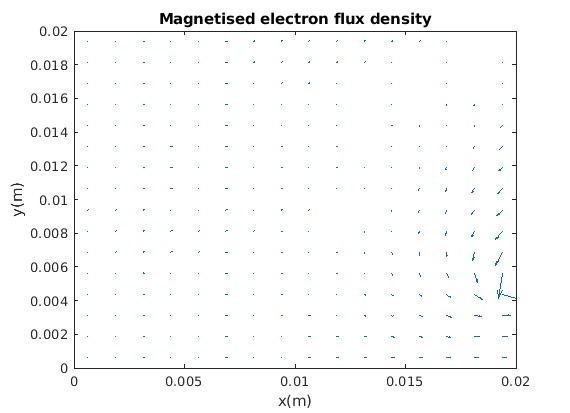}
						\end{tabular}
						\caption{Electron flux density, test case 3, $45^\circ$ magnetic field.} \label{fig.2Dtest2fluxes_45}
					\end{figure}

					Finally, we illustrate in Figures \ref{fig.2Dtest2_45} and \ref{fig.2Dtest2fluxes_45} that our numerical scheme is able to handle test case 3, even with the most extreme magnetic field with $\theta=45^\circ$. This is the most extreme test in the sense that a $45^\circ$ angle maximizes the term $\beta_1\beta_2$ in the magnetic tensor $\mtensor_\elec$ defined in \eqref{def:Mtensor}, which translates to an anisotropic diffusion tensor with very strong cross-diffusion. We now give a detailed discussion about the results in Figures \ref{fig.2Dtest2_45} and \ref{fig.2Dtest2fluxes_45}. We observe in Figure  \ref{fig.2Dtest2fluxes_0}, right  and Figure \ref{fig.2Dtest2fluxes_45}, left that without magnetisation, most of the electrons travel from the left wall to the right wall, parallel to the $x$-axis at the lower part of the domain, and with approximately an angle of  $-45^\circ$ with respect to the $x$-axis at the upper part of the domain. Upon enabling a magnetic field of $\theta=45^\circ$, we see that the flux in $-45^\circ$ direction is greatly decreased (see Figure \ref{fig.2Dtest2fluxes_45}, right), which is expected since the flux is perpendicular to the magnetic field (see Eq.\eqref{eq:mag_field_perp}). We also observe in Figure \ref{fig.2Dtest2_45} that the maximum electron density is located near the center of the wall. This is due to the fact that above this region, there is a loss of electrons towards the upper boundary, whereas below this region, the electrons are attracted to the bottom right.

					\section{Conclusion}\label{sec:Conc}
					\noindent In this work, we propose the use of a combined hybrid mimetic mixed (HMM) method with the Scharfetter-Gummel (SG) scheme for magnetised transport problems encountered in plasma physics. The main reason behind the employment of the HMM method is due to its ability to handle anisotropic diffusion tensors on generic meshes, which is mainly encountered in the presence of magnetic fields. Another advantage of hybrid schemes is that the stencils involved for solving the linear system of equations are purely local, which allows the use of static condensation for a very efficient implementation. Another important factor we introduced is a grid-based P\'eclet number which captures the relative strength of advection over diffusion properly, which was used for the definition of the advective fluxes. 
					
					Numerical tests show that on a standard one-dimensional model, the proposed scheme is slightly better than the Scharfetter-Gummel scheme, both on fine and coarse meshes. It is also important to note that even though the peaks of the solution profiles are not captured properly on the coarse mesh, both schemes still capture the shape (location of density peaks and quasi-neutral regions) of the solution correctly. For two-dimensional models, the numerical results obtained from our proposed scheme agree with what is physically expected, even for distorted meshes and for extreme tests which involve highly anisotropic magnetic fields with strong cross-diffusion terms. On the other hand, it is not straightforward to extend the classical Scharfetter-Gummel scheme in order for it to handle the anisotropic diffusion that arises from the presence of the magnetic field. 
					
					Numerical tests showed the robustness and applicability of the HMM-SG method on both uniform and distorted quadrilateral meshes. Hence, an interesting and natural follow-up would be the application of the proposed method onto the magnetic field aligned meshes in \cite{PGO16-anisotropy,ZPFA19-numerics}. Another interesting aspect to consider would be the application of the proposed scheme to models in cylindrical coordinates, which is commonly encountered in applications. Another avenue that may be explored for future research would involve using the complete flux scheme \cite{LDBMM13-CF,AB11-FVCF} or hybrid high order schemes \cite{DSGD15-adv-diff,DE15-HHO} for the spatial discretisation, which will allow us to resolve the peaks of the solution profile, even on coarse meshes. We also note that due to the use of a first-order Euler time integration scheme for the continuity equations, we might not have been able to fully take advantage of the second-order approximation of the charge densities. Hence it might also be interesting to investigate the use of second-order time integration schemes for the continuity equations, especially if it allows us to preserve the accuracy of the solution whilst taking much larger time steps. These are quite challenging aspects to explore since a proper study and choice of flux limiters would be needed to ensure that the numerical solutions do not exhibit nonphysical oscillations.
					\bibliographystyle{abbrv}
					\bibliography{num_md2d}
				\end{document}